\pdfoutput=1

\documentclass[10pt, conference]{IEEEtran}
\usepackage{cite}
\usepackage{amsmath,amssymb,amsfonts}
\usepackage{algorithmic}
\usepackage{graphicx}
\usepackage{textcomp}
\usepackage{color}
\usepackage{xcolor}
\usepackage{xspace}
\usepackage[hyphens]{url}
\usepackage{fancyhdr}
\usepackage[bookmarks=true,breaklinks=true,letterpaper=true,colorlinks,citecolor=blue,linkcolor=blue,urlcolor=blue]{hyperref}

\usepackage{tikz}
\usepackage{float}
\usepackage{subfig}
\usepackage{multirow}
\usepackage{makecell}

\usepackage{url}

\newcommand{\hpcayear}{2025}

\newcommand{\hpcasubmissionnumber}{428}

\def\hpcacameraready{} 

\newcommand\hpcaauthors{
{\large Yuda An$^1$, Shushu Yi$^1$, Bo Mao$^2$, Qiao Li$^2$, Mingzhe Zhang$^3$} \\
{\large Ke Zhou$^4$, Nong Xiao$^5$, Guangyu Sun$^{1,6}$, Xiaolin Wang$^1$, Yingwei Luo$^1$, Jie Zhang$^1$}
}
\newcommand\hpcaaffiliation{
{\textit{Computer Hardware and System Evolution Laboratory,}} \\
{Peking University$^1$, Xiamen University$^2$,} \\
{Institute of Information Engineering, Chinese Academy of Sciences$^3$,} \\
{Wuhan National Laboratory for Optoelectronics of Huazhong University of Science and Technology$^4$,} \\
{Sun Yat-sen University$^5$,} \\
{Beijing Advanced Innovation Center for Integrated Circuits$^6$} \\
{https://www.chaselab.wiki}
}
\newcommand\hpcaemail{}

\title{A Novel Extensible Simulation Framework for CXL-Enabled Systems} 

\author{
  \ifdefined\hpcacameraready
    \IEEEauthorblockN{\hpcaauthors{}}
      \IEEEauthorblockA{
        \hpcaaffiliation{} \\
        \hpcaemail{}
      }
  \else
    \IEEEauthorblockN{\normalsize{HPCA \hpcayear{} Submission
      \textbf{\#\hpcasubmissionnumber{}}} \\
      \IEEEauthorblockA{
        Confidential Draft \\
        Do NOT Distribute!!
      }
    }
  \fi 
}

\newcommand{\mycomment}[1]{}
\newcommand{\ignore}[1]{}

\newcommand{\eg}{\textit{e.g.}, }
\newcommand{\ie}{\textit{i.e.}, }

\newcommand \footnoteONLYtext[1]
{
	\let \mybackup \thefootnote
	\let \thefootnote \relax
	\footnotetext{#1}
	\let \thefootnote \mybackup
	\let \mybackup \imareallyundefinedcommand
}

\begin{document}
\maketitle

\thispagestyle{plain} 
\pagestyle{plain} 

\newcommand{\hpcaheight}{0mm}
\ifdefined\eaopen
\renewcommand{\hpcaheight}{12mm}
\fi

\begin{abstract}
Compute Express Link (CXL) serves as a rising industry standard, delivering high-speed cache-coherent links to a variety of devices, including host CPUs, computational accelerators, and memory devices. It is designed to promote system scalability, enable peer-to-peer exchanges, and accelerate data transmissions. To achieve these objectives, the most recent CXL protocol has brought forth several innovative features, such as port-focused routing, device-handled coherence, and PCIe 6.0 compatibility. However, due to the limited availability of hardware prototypes and simulators compatible with CXL, earlier CXL research has largely depended on emulating CXL devices using remote NUMA nodes. Unfortunately, these NUMA-based emulators have difficulties in accurately representing the new features due to fundamental differences in hardware and protocols. Moreover, the absence of support for non-tree topology and PCIe links makes it complex to merely adapt existing simulators for CXL simulation. To overcome these problems, we introduce ESF, a simulation framework specifically designed for CXL systems. ESF has been developed to accurately reflect the unique features of the latest CXL protocol from the ground up. It uses a specialized interconnect layer to facilitate connections within a wide range of system topologies and also includes key components to carry out specific functions required by these features. By utilizing ESF, we thoroughly investigate various aspects of CXL systems, including system topology, device-handled coherence, and the effects of PCIe characteristics, leading to important findings that can guide the creation of high-performance CXL systems. The ESF source codes are fully open-source and can be accessed at \url{https://anonymous.4open.science/r/ESF-1CE3}.
\end{abstract}

\section{Introduction}
\label{sec:intro}
With the prevalence of large-scale data-intensive applications such as artificial intelligence, life science, and climate modelling~\cite{(MLexample)Li2015HeteroSparkAH,(GPT-3)Brown2020LanguageMA,(gpt4)OpenAI2023GPT4TR,(life-sci1)Langmead2009UltrafastAM,(life-sci2)Chowdhury2020ADR,(life-sci3)Angizi2019AlignSAP,(climate1)Giorgi2019ThirtyYO,(climate2)Yuval2020StableMP,(climate3)Nguyen2023ClimateLearnBM}, there are increasing demands to aggregate tons of computation and memory resources into a uniform system. Peripheral component interconnect express (PCIe)~\cite{(PCIe5.0),(PCIe6.0)}, as one of the most popular interconnect standards, has been widely adopted in the computing system to connect between the host CPU and diverse peripheral devices including graph processing units (GPUs) and solid-state drives (SSDs)~\cite{(GPU1)Zhang2020ZnGAG,(GPU2)Hong2024BandwidthEffectiveDC,(SSD1)Tavakkol2018MQSimAF,(SSD2)Jung2022HelloBB}. 

Compared to other types of interconnects (e.g., Ethernet~\cite{(eth)}, SATA~\cite{(sata)}, and DDR~\cite{(ddr)}), PCIe can deliver much higher aggregated throughput (\eg 256 GB/s in 16 PCIe 6.0 lanes~\cite{(PCIe6.0)}). In addition, PCIe supports various communication protocols (\eg NVMe~\cite{(nvme)}), exhibiting high compatibility. However, PCIe fails to extend the host local memory with external PCIe memory devices due to the lack of coherence mechanisms~\cite{(SSD2)Jung2022HelloBB}. Specifically, the memory accesses that target PCIe device memory address space are required to be \emph{non-cachable}. CPU cores must directly access the PCIe device memory and are not allowed to store copies of data from the device memory within their internal caches. Software involvement is necessary to maintain data coherence. This limitation significantly worsens the memory access performance. Thus, building computation and memory pools atop PCIe cannot satisfy the demands of large-scale data-intensive applications.

Compute Express Link (CXL) is an emerging industry standard that offers high-performance cache-coherent interconnect capability to heterogeneous devices, including host CPUs, computational accelerators, and memory devices~\cite{(CXL3.1),(CXL3.1WP)}. CXL is designed to operate over the existing PCIe infrastructure, which utilizes the same physical and electrical interfaces. This design philosophy aids CXL with the high performance and backward compatibility of PCIe technology. CXL also provides the features of cache coherency and memory semantic support, which can seamlessly extend the host-side processor and memory with the external CXL accelerators and memory devices. Thus, CXL enables efficient data sharing and communication within computation and memory pools. 

While CXL has great potential to change the existing computer architecture, most of the prior studies on CXL~\cite{(Pond)Li2022PondCM,(TPP)Maruf2022TPPTP,(priorEmu)Arif2022ExploitingCM} leverage remote NUMA nodes to emulate CXL devices due to the lack of hardware prototypes. As high-version CXL is still at the proof of concept (PoC) stage, we believe constructing a CXL simulator would be the wheel to drive the high-performance interconnect research forward. Nevertheless, it is challenging to simply extend the existing simulators and emulators to support the CXL simulation. Specifically, the CXL standard aims to achieve ultra-high scalability by providing complicated non-tree system topology and coherent peer-to-peer communication. However, NUMA-based emulators, which have been adopted in previous work, face strict physical constraints (\eg socket number) and fail to extend system scalability. 
Prior computation-centric simulators, such as gem5~\cite{(gem5),(gem5-dram)} and GPGPUsim \cite{(GPGPUsim)Bakhoda2009AnalyzingCW}, focus on accurate processor unit modeling, however, support only legacy interconnects (\eg gem5 only supports legacy PCI links), which are unable to operate in non-tree topologies.
On the other hand, network-centric simulators, such as BookSim \cite{(BookSim)Jiang2013ADA} and Garnet \cite{(garnet)Agarwal2009GARNETAD}, pay attention to diverse network topologies and flow control mechanisms. These simulators lack the support of coherency management, which is considered a key promise of the CXL standard. In summary, existing tools struggle to reflect critical features of the CXL standard.

Tackling the aforementioned challenges, we propose our novel extensible simulation framework, \emph{ESF}, that is built atop the CXL backbone. This framework introduces two function layers for an accurate simulation of highly scalable CXL systems, namely \emph{interconnect layer} and \emph{device layer}. The interconnect layer is dedicated to supporting complicated system topologies. Upon system initialization, this layer constructs a topology graph of the system and provides detailed routing information to the devices for intercommunication uses. On the other hand, the device layer models several types of fundamental CXL devices, including CXL accelerators, memory devices, and CXL switches. During simulation, these devices conduct CXL protocol functions and communicate with each other by leveraging the communication function of the interconnect layer. For example, CXL switches build internal routing tables based on the topology information provided by the interconnect layer and route different requests to the correct destinations. The tight collaboration of these two layers ensures ESF to accurately simulate a highly scalable system defined by the CXL standard. The validation experiment proves the accuracy of ESF with errors ranging from 0.1\% to 10\%. With accurate simulation, ESF can uncover several issues that the existing simulators are unable to figure out, including the performance impacts of diverse system topologies and the design choices for device-managed coherence. 

Our \textbf{contributions} are summarized as follows:

\noindent $\bullet$ \emph{CXL simulation challenge analysis of existing research tools:}
The CXL standard is aimed at supporting rack-level systems with scalable performance, which requires complicated non-tree system topology and coherent peer-to-peer communication. To meet these requirements, the CXL protocol introduces several novel features,
including port-based routing, device-managed coherence and the adoption of high-version PCIe physical links.
Unfortunately, existing simulation and emulation tools face challenges in accurately reflecting these critical features. Most of the prior works adopt remote NUMA nodes as CXL hardware emulators. However, the physical limitation of NUMA platforms prevents them from emulating port-based routing. Meanwhile, existing computation-centric simulators lack the support of PCIe simulation, while network-centric simulators fail to provide coherence management functionality. 

\noindent $\bullet$ \emph{Novel simulation framework customized for CXL systems:} 
To address the challenges in existing tools comprehensively, we propose a customized simulation framework, ESF, which consists of two fundamental layers, namely the interconnect layer and the device layer. While the interconnect layer is dedicated to providing interconnection and scalability of the simulated system, the device layer performs device-specific functions, such as coherence management. The novel framework carefully implements a set of components to model the essential features of CXL. Firstly, it provides a switch component that supports PBR. Secondly, it implements a device-side inclusive snoop filter as an example of device coherency agent (DCOH). Lastly, it implements the bus components while considering unique characteristics of PCIe buses to accurately reflect the behaviors of real CXL platforms.


\noindent $\bullet$ \emph{Exploration on the performance impacts of multiple new CXL features:} 
We perform a set of experiments to explore the performance impacts of emerging CXL features in multiple representative systems implemented with our novel simulation framework. Our investigation focuses on three main aspects: (1) the impacts of different system topologies, (2) the impacts of device-managed coherence, and (3) the unique full-duplex feature of PCIe transmission. From the experimental results, we derive three key observations. First, the traditional tree-like system topology experiences severe bandwidth and latency bottlenecks at the root, leading to potential performance degradation similar to systems with a chain-like topology. Second, the device-side inclusive snoop filter receives unique request patterns because most of the requests that reach the snoop filter are cache misses. Therefore, a customized structure is essential for the snoop filter to achieve optimal performance. Third, we observe from read-write mixed workloads that full-duplex transmission of PCIe buses results in a bandwidth improvement compared to those with a single type of access pattern. These observations pave the road to future CXL system designs.

\section{Background, Motivation and Challenges}
\label{sec:background}
\subsection{Basic Features in CXL Protocol}
Compute Express Link (CXL) is an emerging standard that provides high-performance and cache-coherent interconnects for heterogeneous devices ranging from host CPUs to memory devices~\cite{(CXL3.1), (CXL3.1WP)}. To this end, the CXL standard introduces various new features. In the following, we will elaborate on these features in detail.

\begin{figure}
    \centering
    \includegraphics[width=1\linewidth]{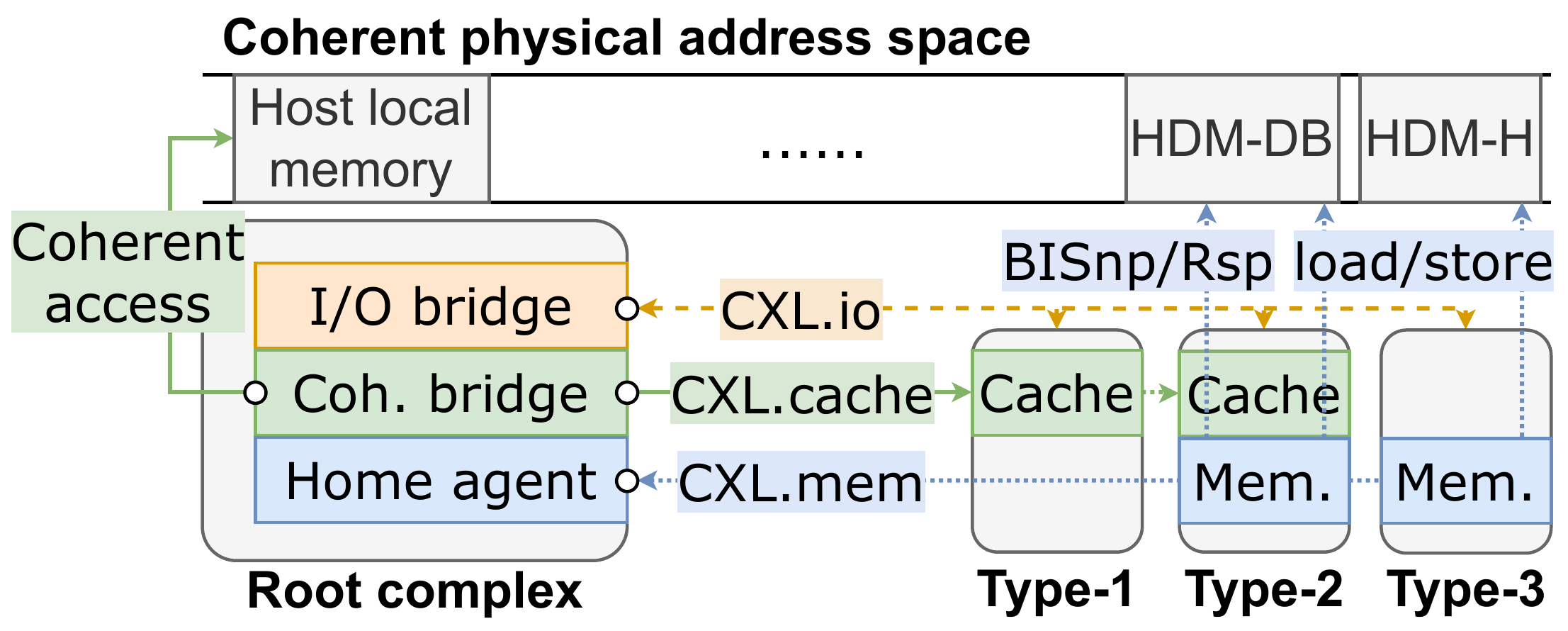}
    \vspace{-15pt}
    \caption{CXL sub-protocols, endpoint types, and root complex.}
    \vspace{-20pt}
    \label{fig:bg_protocol}
\end{figure}

\noindent \textbf{CXL sub-protocols and Flex Bus layers.}
CXL provides both backward-compatible and incremental functions over PCIe physical and electrical interfaces via three sub-protocols: \emph{CXL.io}, \emph{CXL.cache} and \emph{CXL.mem}, as depicted in Figure~\ref{fig:bg_protocol}.

The CXL.io protocol, highlighted in dashed yellow lines in the figure, is the fundamental protocol that all CXL devices and host CPUs need to support. It is responsible for all basic I/O operations, including PCIe backward-compatible operations, device enumeration, and device configuration. CXL.io mainly adopts the command set of traditional PCIe links with slight enhancements. In contrast, the other two protocols, CXL.cache and CXL.mem, employ customized command sets.

A traditional PCIe device counts on DMA mechanism~\cite{(dma1)Su2011APM,(dma2)Rhu2017CompressingDE} to access data that reside in host memory. This mechanism brings two drawbacks. First, DMA is optimized for massive contiguous accesses and shows suboptimal performance when servicing small accesses. Second, DMA does not provide data coherence guarantee, which necessitates software assistance, limiting the access speed. To address this challenge, CXL proposes CXL.cache protocol, as highlighted in solid green lines in the figure, which enables cacheline-grained coherent access from devices to the host memory via hardware assistance, eliminating software involvement.

The CXL.mem protocol exposes the CXL device internal memory to the host as coherent memory space (cf. dotted blue lines in Figure~\ref{fig:bg_protocol}). It provides byte-addressable memory semantics, allowing the host to access the device local memory via \texttt{load/store} instructions. The CXL.mem enables the expansion of coherent physical memory through PCIe ports, constructing a highly scalable memory system.

\begin{figure}[t]
    \centering
    \includegraphics[width=1\linewidth]{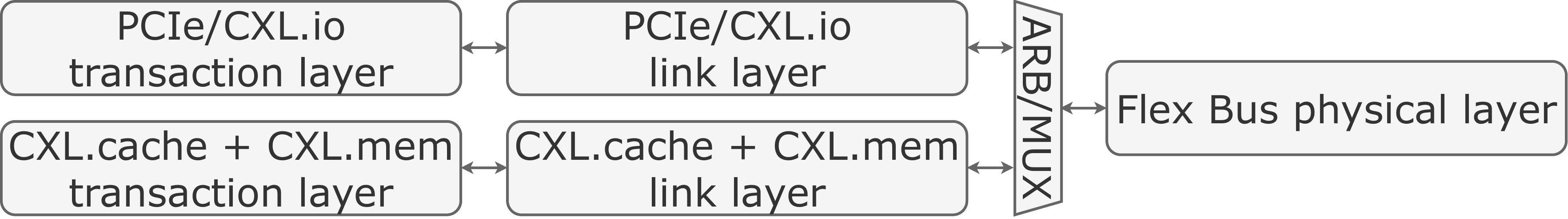}
    \caption{Hierarchy of CXL Flex Bus layers.}
    \vspace{-20pt}
    \label{fig:bg_flex}
\end{figure}

To achieve fast and reliable transmission, PCIe protocol consist of three layers, namely \emph{transaction} layer, \emph{link} layer, and \emph{physical} layer~\cite{(PCIe5.0),(PCIe6.0)}. CXL, built on top of PCIe, adopts these layers. Moreover, as only the CXL.io protocol, rather than CXL.cache and CXL.mem, operates on a PCIe backward-compatible command set, CXL extends the transaction and link layers for CXL.cache and CXL.mem. The entire hierarchy is called \emph{Flex Bus}~\cite{(CXL3.1)}, depicted in Figure~\ref{fig:bg_flex}. The transaction layer is where requests are handled by different CXL protocols. The link layer transforms one or more requests into packets and is responsible for transmission reliability. 
After the packets are ready, the physical layer performs the actual transmission. Note that the physical layer is shared by all protocols. Therefore, an arbitrator/multiplexer (CXL ARB/MUX) between the physical layer and the link layer is required to prepare the physical bus for a specific protocol and distributes a physical packet to the correct link-layer handler.

\noindent \textbf{CXL endpoint types and root complex.}
CXL enables the integration of various types of peripheral devices in the computing system. These devices are distinct from the hosts and named as \emph{Endpoints} (EP). On the contrary, the various components on the host side that conduct CXL functions are named \emph{Root Complex} (RC). As shown in Figure~\ref{fig:bg_protocol}, the endpoints are categorized into three types. As a backward compatibility support, all the types of devices support CXL.io protocol for I/O operations. In the root complex, the CXL.io protocol is managed by the I/O bridge. In addition to I/Os, different types of CXL devices target different functions.

Type-1 devices are endpoints with a fully-coherent cache but without a global-visible device local memory. The cache buffers data using the CXL.cache protocol from the host-side memory to utilize the potential data locality. 
Devices that do not expose their local memory, such as SmartNIC~\cite{(SNIC)Firestone2018AzureAN}, match with type-1.
In the root complex, the CXL.cache protocol is mainly served by the coherency bridge. It responds to CXL.cache requests from devices and records coherence metadata of cached data. It may also actively send requests to devices when the host asks for ownership or copies of the cached data.

A type-2 device contains local memory components (\eg DDR DRAM or HBM modules) in addition to a coherent local cache. 
Traditional computer architecture cannot coherently access the memory of peripheral devices. As a result, it relies on explicit data migration between host and device to feed the computational cores in PCIe-attached type-2 devices, causing underutilized performance~\cite{(GPUmemWall1)Rajbhandari2021ZeROInfinityBT,(GPUmemWall2)Zhang2023G10EA,(GPUmemWall3)Zhou2023MPressDB}.
To address this issue, the CXL protocol provides methods for the host and devices to efficiently communicate with each other. Specifically, the host can push operands to and pull results from the device memory via CXL.mem protocol with hardware assistance, and the devices can directly access data in host memory through CXL.cache protocol without waiting on explicit data copy. 

To support this intercommunication, a CXL type-2 device needs to expose its local memory address space to the entire system. This address space is integrated into the host coherent physical address space and is named \emph{Host-managed Device Memory} (HDM). Figure~\ref{fig:bg_protocol} shows the integration of HDM, with the host local memory and HDM regions organized in the same physical address space. 
In the root complex, in addition to the I/O bridge, two components are involved in communication with type-2 devices. The coherency bridge serves the CXL.cache requests, similar to the case of type-1 devices. The home agent handles CXL.mem transactions, such as issuing load/store instructions to type-2 devices.

Similar to type-2 devices, the CXL type-3 devices also include local memory components that are exposed as fully-coherent HDMs. However, typical type-3 devices do not contain computational cores or coherent local caches. These devices are categorized as memory expanders~\cite{(SMT)Kim2023SMTSM}, which extend the total memory capacity of the entire system. In the root complex, the I/O bridge and the home agent manage CXL.io and CXL.mem transactions to type-3 devices, respectively.

\noindent \textbf{HDM coherence management modes.}
Type-2 and type-3 devices expose their local memory as HDMs, which are required to be fully coherent. CXL proposes three modes to manage the coherence of HDMs. The first mode is host-managed coherence, denoted as HDM-H. When an HDM is in HDM-H mode, the host is fully responsible for managing its coherence (based on either software or hardware), and the device is not required to act for coherency. 
The second mode is called device-managed coherence using \emph{Back-Invalidate Snoop} (BISnp), denoted as HDM-DB. In HDM-DB mode, the device is responsible for managing the coherence of its local memory and can actively send BISnp requests to other devices, including the host. A BISnp request may ask the device for a cacheline that has been cached previously. Upon receiving a BISnp request, the device needs to check whether it has cached the corresponding cacheline, and needs to flush it back through \emph{Back-Invalidate Response} (BIRsp). 
There are two critical facts in the latest CXL 3.1 version about HDM-DB mode. First, the BISnp/Rsp transactions are sent through CXL.mem protocol, but not through CXL.cache. The CXL.mem provides two dedicated channels used only for BISnp and BIRsp. Second, to utilize the 64GT/s PCIe 6.0 transmission speed, type-2/3 devices that support device-managed coherence must operate in HDM-DB mode. This means that the HDM-DB will be the main mode used to achieve optimal performance with device-managed coherence. 
The third mode is called ``device coherent" and is denoted as HDM-D. In the latest version of CXL, this mode is left for backward compatibility for devices that manage HDM coherence through the CXL.cache protocol. In the rest of this paper, we mainly consider HDM-DB as the typical mode of \emph{Device-Managed Coherence} (DMC). 

\subsection{The Scale-up of CXL-Enabled Systems\label{sec:2B}}
To fulfill the growing computational needs of emerging large-scale applications (\eg machine learning~\cite{(MLexample)Li2015HeteroSparkAH,(GPT-3)Brown2020LanguageMA,(gpt4)OpenAI2023GPT4TR}, life science~\cite{(life-sci1)Langmead2009UltrafastAM,(life-sci2)Chowdhury2020ADR,(life-sci3)Angizi2019AlignSAP} and climate modelling~\cite{(climate1)Giorgi2019ThirtyYO,(climate2)Yuval2020StableMP,(climate3)Nguyen2023ClimateLearnBM}), CXL aims to support the scale-up of a CXL-enabled system from a single node to rack-level and even further to extend the memory capacity and computation capability. To this end, the CXL protocol projects the design of CXL switch as the key component, which is responsible for large-scale interconnection. It introduces port-based routing (PBR) and multi-level switching as the key features, which distinguish CXL switches from traditional PCIe switches~\cite{(PCIe5.0),(PCIe6.0)}. In this part, we briefly introduce the novel CXL switches.

\begin{figure}
    \centering
    \includegraphics[width=1\linewidth]{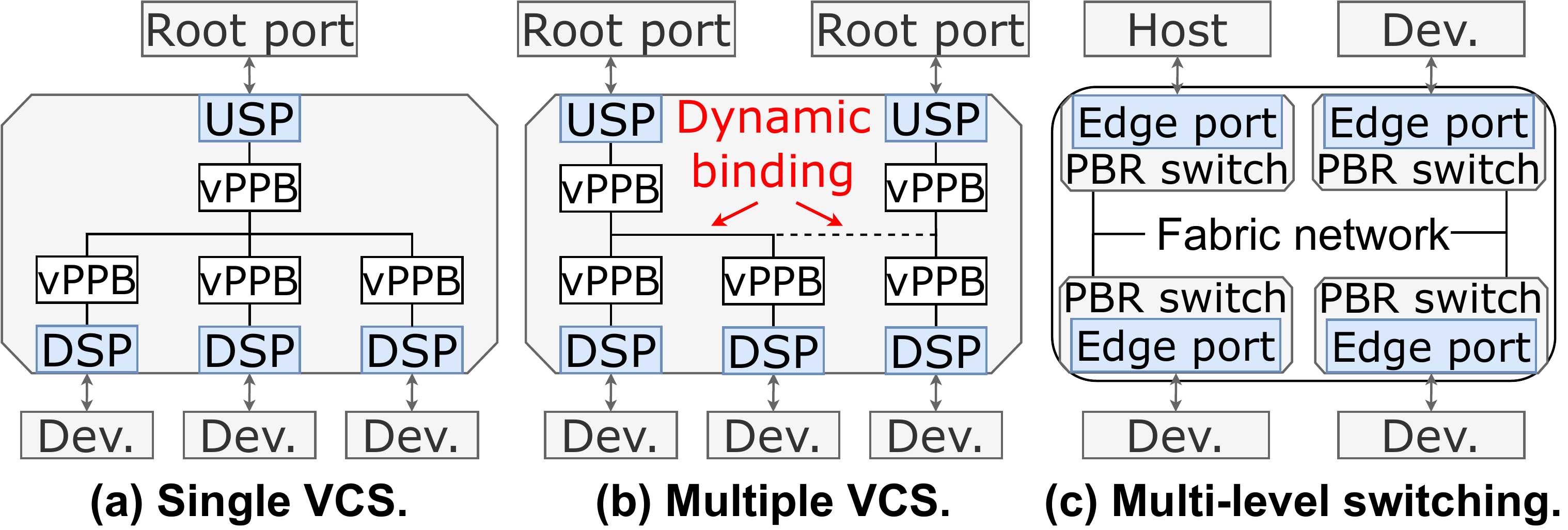}
    \vspace{-15pt}
    \caption{CXL system switching configuration examples.}
    \vspace{-20pt}
    \label{fig:bg_switch}
\end{figure}

\noindent \textbf{PCIe-compatible CXL switch.}
CXL switches can operate in a PCIe-compatible configuration. The basic PCIe-compatible configuration of a CXL switch is known as a \emph{Single Virtual CXL Switch} (Single VCS), which is depicted in Figure~\ref{fig:bg_switch}a. A Single VCS consists of a single \emph{Upstream Port} (USP) and one or more \emph{Downstream Ports} (DSP). Similar to traditional PCIe switches, the ports are connected via \emph{virtual PCI-to-PCI Bridges} (vPPB). The USP of a Single VCS links to a root port, leading to a host or another switch. Each DSP of a Single VCS links to a CXL or legacy PCIe device including another switch. A Single VCS behaves identically to a PCIe switch except that it supports CXL protocols at the link and transaction layers. 

Figure~\ref{fig:bg_switch}b shows another PCIe-compatible configuration, namely the \emph{Multiple VCS}. This configuration supports multiple USPs, each linked to a root port, allowing multiple \emph{requesters} (hosts and accelerators) to issue downstream requests. The multiple VCS offers additional features compared to the Single VCS. During initialization, a Multiple VCS presents itself as multiple Single VCS instances, corresponding to the number of USPs. The association between DSPs and USPs can be dynamically configured or even software-composed during execution. Additionally, multiple vPPBs can be combined into a single physical port, thereby exposing the physical device under this physical port as multiple logical devices, providing resource isolation and pooling of a single CXL device.

\noindent \textbf{Multi-level switching.}
Although multiple VCSs offer some innovative features, fundamentally, they still operate in a PCIe-compatible manner. A key limitation is that each logical CXL device under a DSP is linked to a single USP. Peer-to-peer communication without host assistance is not supported. Thus, the scalability of the entire system is limited. To address this challenge, the CXL protocol proposes \emph{multi-level switching} on top of \emph{Port-Based Routing} (PBR). A system that supports multi-level switching mainly consists of two parts: the \emph{fabric network} and the peripheral components, depicted in Figure~\ref{fig:bg_switch}c. The fabric network is constructed from multiple PBR switches. Ports connecting to devices/hosts that are not PBR switches (\ie on the edge of the fabric network) are called \emph{edge ports}. PBR is used to route requests between different edge ports. 
In a multi-level switching system, each edge port is assigned a 12-bit port ID, supporting up to 4096 edge ports. 
When a CXL.mem request arrives at the edge port, the PBR switches route it to the correct edge port based on its internal routing table. 
This mechanism enables peer-to-peer communication among CXL devices and allows diverse non-tree system topologies in the fabric network. The design philosophy of such topology flexibility is aimed to enhance overall system performance.

\subsection{The Necessity of CXL Simulator}
There exist multiple challenges that necessitate a CXL simulator to conduct relevant computer architecture research.

\noindent \textbf{The lack of hardware prototypes supporting the latest CXL specification.} 
Many of the aforementioned features of CXL are proposed in the latest CXL 3.1 specification, including PCIe 6.0 64GT/s transmission speed, multi-level switching, and coherent peer-to-peer communication support. These features are critical to establishing a CXL system with high scalability and enhanced performance. However, none of the current CXL hardware prototypes~\cite{(Xeon),(CXLMXC),(CXL-FPGA),(Micron-CXL)} are compatible with CXL 3.1. This lack of hardware has greatly hindered research progress in the field of CXL. A software simulator can help researchers study the newest features of CXL standards, boosting the development of high-performance design of CXL systems. 

\noindent \textbf{CXL researches demand a highly configurable multi-level system.}
As an interconnect standard, different parts in a CXL system tightly work together with each other. Thus, CXL research involves multiple aspects including the specification, the devices, and the whole system. Simply evaluating an individual aspect is inefficient for studying the overall system behaviors. Moreover, building a high-performance CXL system requires tuning the system configuration by taking all these aspects into consideration. However, hardware implementation costs tremendous labor and time efforts. Leveraging a CXL simulator, researchers can easily test a wide range of system setups without suffering from implementing specific hardware.

\noindent \textbf{Existing tools struggle to simulate CXL systems.} 
Due to the lack of hardware, most of the prior works leverage an emulation methodology to study CXL systems~\cite{(Pond)Li2022PondCM,(TPP)Maruf2022TPPTP,(priorEmu)Arif2022ExploitingCM}. This methodology treats remote NUMA nodes as an emulator of CXL devices. However, as reported in~\cite{(demystify)Sun2023DemystifyingCM}, there are critical differences between remote NUMA nodes and real CXL devices. With the new features proposed in the CXL 3.1 specification, this gap of behavior and performance between NUMA-based emulators and CXL devices will become more severe, making NUMA-based emulators an inaccurate choice for extensive studies. 
In addition to NUMA-based emulators, there exist plenty of software simulators for computer architecture. These simulators can be categorized in two types: (1) computation-centric simulators, including gem5~\cite{(gem5),(gem5-dram)} and GPGPUsim~\cite{(GPGPUsim)Bakhoda2009AnalyzingCW}, and (2) network-centric simulators, including BookSim~\cite{(BookSim)Jiang2013ADA} and Garnet~\cite{(garnet)Agarwal2009GARNETAD}. Different types of simulators focus on different aspects of the system. The computation-centric simulators focus on accurate processor unit modeling, while the network-centric simulators pay attention to diverse network topologies and flow control. However, since CXL systems are highly collaborative systems, in which all the aspects play an important role in system performance, these simulators struggle to provide comprehensive simulation results. 

Specifically, there are three major characteristics of CXL-enabled systems that existing tools are unable to accurately simulate. First, CXL 3.1 specification introduces PBR and multi-level switching to support high system scalability and non-tree topologies. Second, the specification defines DMC as a key feature to support direct peer-to-peer communication between devices without the assistance of the host. Devices featuring DMC functionality are required to have a \emph{Device COHerency agent} (DCOH) to manage coherence for the cachelines from their local memory. With the help of DMC devices, the system can offload expensive coherence management to multiple individual devices, eliminating the need for a central coherence engine and thereby enhancing performance scalability. Third, the adopted PCIe bus and its full-duplex nature notably influence system performance. None of the existing tools can accurately simulate all these features. NUMA-based emulators, being limited by physical constraints such as the number of sockets and slots, can hardly provide the required scalability. They are unable to simulate either DMC or PCIe bus, because they can only adopt a traditional memory system with DDR DRAM that doesn't support DMC, and the data paths of NUMA interfaces are quite different from CXL-enabled PCIe paths. Existing computational simulators can only be configured to basic switch setups and often leverage a central engine (e.g., host-side CPU) for coherence management. Thus, they struggle to simulate PBR and DMC. On the other hand, existing network-centric simulators lack the basic support of coherence management. In addition, existing simulators provide poor support for high-version PCIe simulation (\eg gem5 only supports legacy PCI simulation). To sum up, the existing simulation tools face intractable difficulties in accurately simulating CXL systems. This lack of tools urges the development of a CXL simulator.

\section{Modelling Details}
\label{sec:model}
\begin{figure}
    \centering
    \includegraphics[width=1\linewidth]{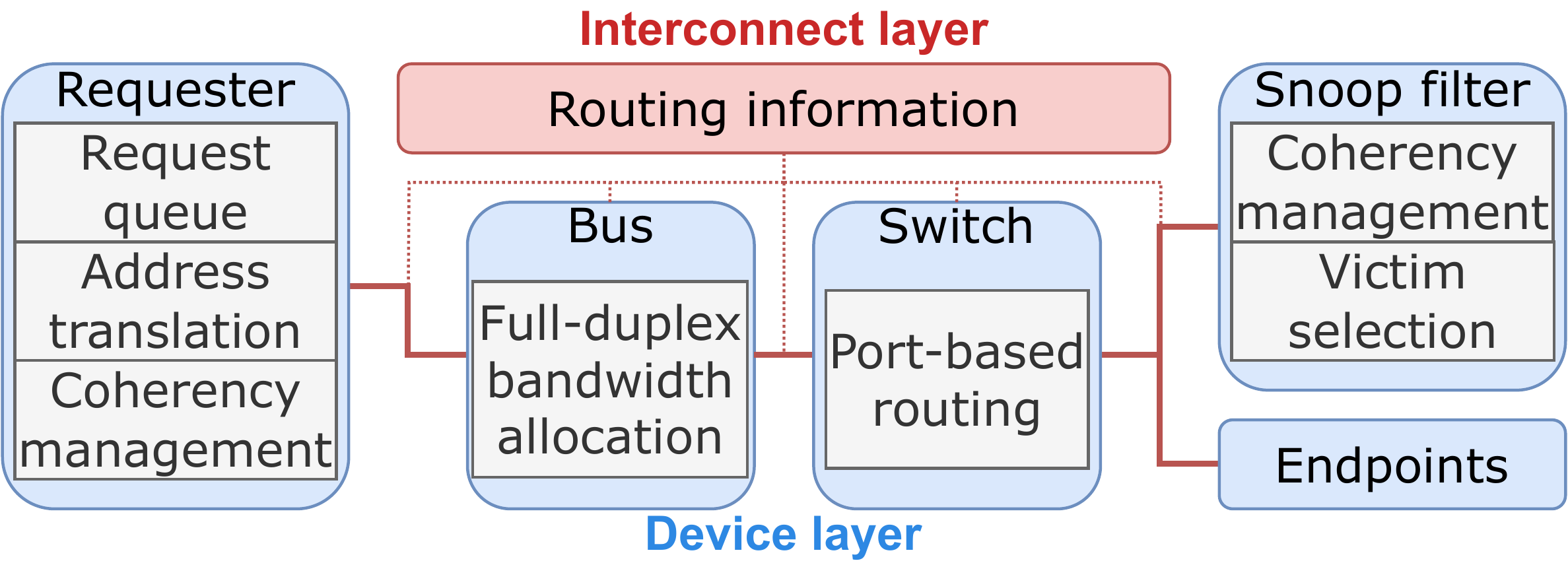}
    \vspace{-20pt}
    \caption{Overview of the ESF architecture and components.}
    \vspace{-20pt}
    \label{fig:arch}
\end{figure}

In this work, we present, \emph{ESF}, our novel simulation framework developed from scratch to support accurate simulation of the aforementioned critical features. Figure~\ref{fig:arch} depicts the architectural overview of ESF and its major components. In the rest of this section, we introduce the design principles of ESF and describe these major components.

\subsection{Architectural Overview}
To support the simulation of highly scalable CXL-based systems, our novel simulation framework utilizes a dedicated architecture consisting of two major layers, which tightly work together to provide accurate and detailed results during the simulation. 

The first layer is the interconnect layer. As mentioned, to provide high scalability, the CXL protocol proposes multi-level switching and port-based routing. These features allow CXL systems to adopt non-tree interconnect topologies that most of the existing tools cannot support. The interconnect layer in ESF is dedicated to the simulation of diverse system topologies. Upon initialization, the interconnect layer receives a set of device pairs, which are configured as directly connected through physical links. Then, the interconnect layer constructs an internal topology graph and builds a default routing strategy based on the shortest-path algorithm~\cite{(shortpath1)FagenUlmschneider2008ShortestP,(shortpath2)Sommer2014ShortestpathQI}. During the simulation, the interconnect layer provides routing information to all devices. While most of the devices can directly employ the default routing, devices like switches can access detailed graph information to create dedicated routing for their special functionalities. Therefore, the interconnect layer enables peer-to-peer communication between different devices in a system with an arbitrary topology. 

The next layer on top of the interconnect layer is the device layer. This layer models several kinds of fundamental devices in a CXL system, such as memory devices and physical buses. To fully support peer-to-peer communication as required by the CXL standard, all the devices are treated equally. They can actively operate without involving any central device, such as a host CPU. This enables ESF to simulate key features proposed by CXL protocol, including device-managed coherence (DMC). In addition, the device layer adopts a decoupling design to model device functions. For example, the device coherency agent (DCOH), which mainly conducts DMC function, is decoupled from the memory device. Therefore, ESF can easily configure a wide range of hardware setups, adjusting parameters such as coherence management policies for each DCOH without bothering the memory devices. This function decoupling enhances the flexibility of the simulation configuration, expanding the exploration spaces. 
Based on the fundamental architecture, ESF implements several primary components to support the simulation of multiple classical scenarios in CXL-based systems. 

\textbf{Implementation and usage details.} ESF is primarily written in C++. By default, users can simply prepare configuration files and pass them to the simulator to setup and simulate a proposed system. Furthermore, ESF provides essential abstraction and interfaces in both the interconnect and device layers, which allow users to easily hack it and implement components for their own purposes. The detailed usage instructions are included in the open-sourced code repository.

\subsection{Computational Components}
ESF provides a straightforward abstraction of computational components (\ie requesters) in CXL systems, namely \emph{hosts} and \emph{accelerators}. Each computational component consists of three primary units: \emph{request queue}, \emph{address translation unit}, and \emph{cache coherence management unit}. A request queue is defined by the queue capacity and the time interval between issued requests. It models the capability of the computational component to issue requests to other devices. 
An address translation unit simulates various interleaving policies and can be used to investigate the impacts of different policies on system performance. It can adjust the strategy of interleaving requests among multiple memory endpoints to improve system bandwidth~\cite{(interleave1)Yang2019AnEG,(interleave2)Lee2023TCATDC,(CXL3.1)}. The cache coherence management unit simulates an internal cache, which records the metadata (\eg source endpoints) of fetched cachelines during simulation. Upon receiving a coherent request (\eg BISnp) from any endpoint, the unit searches the cache for the corresponding cacheline and flushes it back if necessary. This unit collaborates with DCOH to execute device-managed coherence functions. The computational component supports the simulation of various access patterns. It can be configured with a stream pattern or random pattern and will automatically generate requests during the simulation. It can also be set in trace-based mode, which receives external trace files and replays the recorded requests. This mode helps conduct simulations of real-world workloads. 
It is important to note that the implementation of computational components is offered as an easy-to-use default configuration.

\subsection{Interconnective components}
ESF implements two main components that are responsible for the simulation of interconnection: \emph{bus} and \emph{switch}. 
They model interconnect features in the CXL specification (\ie the full-duplex PCIe bus transfer and the port-based routing of CXL switch) that allow non-tree system topologies.
To accurately reflect the full-duplex feature of PCIe buses, ESF implements a bandwidth allocation module for the bus component. First, with the help of the interconnect layer, the bus detects all the data transfer directions that pass through it. Then, to simulate full-duplex functionality, the bus allocates full bandwidth for each direction (cf. the bandwidth allocation unit shown in Figure~\ref{fig:arch}). 
ESF also ensures the bus component is highly configurable. The bandwidth can be configured during initialization. Additionally, the bus can be set to half-duplex with configurable turnaround overheads. This flexibility enables researchers to explore multiple hardware setups.
While the bus component mainly operates in two directions (\ie upstream and downstream), the switch component is used for conducting more complicated port-based routing as required by the CXL specification. During the initialization, the switch can receive multiple connections from different devices up to its number of ports. Then, with the help of routing information provided by the interconnect layer, the switch constructs an internal routing table for different sources and destinations. Upon the arrival of a packet, based on the source, receiving port, and destination, the switch forwards it to the corresponding port according to the routing table. 

\subsection{Device-side snoop filter}
To support DMC, We implements a device-side snoop filter as an example DCOH. As required by the CXL specification, the snoop filter operates in inclusive mode. An inclusive snoop filter is a buffer that records all the cachelines from its corresponding endpoints that are cached by other devices, such as the host and accelerators. Each entry in the buffer stores the coherence metadata of a cacheline, including the coherence state and the owner list. When a new coherent request is received (\eg a host requires exclusive ownership of a cacheline), the snoop filter allocates a new entry to record the updated metadata. In cases of conflicts with other devices that have already owned the cacheline, the snoop filter sends BISnp requests to the original owners before proceeding with the new request. Also, when the buffer is run out of new entry, the snoop filter selects a victim entry and sends the corresponding BISnp requests to clear the entry before serving the new request. To support the above functions, we implements the snoop filter as a fully-associative buffer of its corresponding endpoints. It performs entry allocation, metadata update, BISnp requesting, and victim selection for coherent requests targeting these endpoints. When it is necessary to clear an entry (\ie conflict or victim eviction), the snoop filter sends BISnp requests to the original owners using the default routing strategy provided by the interconnect layer. Once all the BIRsps are collected, the snoop filter clears the entry for the next request. It may also write back the cacheline to the corresponding endpoint if the cacheline is flushed in a dirty state. We also modularizes the victim selection to allow the researchers to evaluate various policies. 

\begin{figure}
\begin{minipage}{1\linewidth}
    \centering
    \includegraphics[width=1\linewidth]{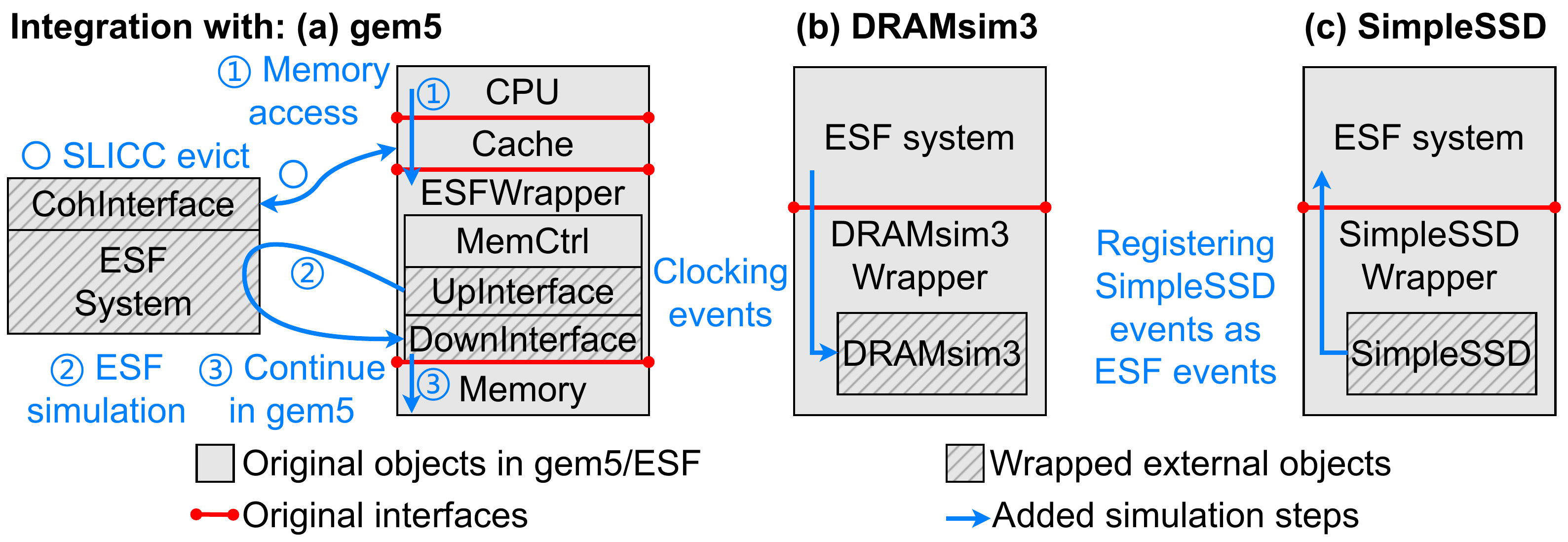}
    \vspace{-15pt}
    \caption{Overview of the integration with other simulators.}
    \vspace{-0pt}
    \label{fig:integrate}
\end{minipage}
\begin{minipage}{1\linewidth}
    \centering
    \resizebox{0.97\linewidth}{!}{
    \begin{tabular}{|c|c|c|}
        \hline
        Supported simulators & Features & Simulated components  \\ \hline
        gem5                 & Event    & Processor micro-architecture  \\
        DRAMsim3             & Cycle    & DRAM endpoint (DDRx, HBM, etc.)  \\
        SimpleSSD            & Event    & SSD endpoint  \\ \hline
    \end{tabular}}
    \vspace{-5pt}
    \captionof{table}{List of simulators integrated with ESF.}
    \vspace{-10pt}
    \label{tab:integrate}
\end{minipage}
\end{figure}


\begin{table}[t]
    \centering
        \begin{tabular}{|c|c|c|}
             \hline
             Simulation frameworks & {\makecell[c]{Off-chip interconnect\\simulation}} & Memory simulation \\ \hline
             ESF & \checkmark & \checkmark \\
             gem5-garnet &  & \checkmark \\
             BZSim &  & \checkmark \\
             CXLMemSim & \checkmark & \checkmark \\
             MQSim\_CXL &  &  \\
             \hline
             \hline
             simulation frameworks & SSD simulation & gem5 integration \\ \hline
             ESF & \checkmark & \checkmark \\
             gem5-garnet &  & \checkmark \\ 
             BZSim &  &  \\ 
             CXLMemSim &  &  \\
             MQSim\_CXL & \checkmark &  \\ \hline 
        \end{tabular}
    \vspace{-5pt}
    \caption{Comparison with other simulation frameworks.}
    \vspace{-20pt}
    \label{tab:othersim}
\end{table}

\subsection{Integration with existing simulators}
In order to demonstrate the extensibility of our framework, we integrate it with several existing simulators. Table~\ref{tab:integrate} provides an overview of these simulators. The first integrated simulator is \emph{gem5}~\cite{(gem5), (gem5-dram)}. As a widely adopted simulator, gem5 models processors and memory systems with extensive details. We integrate ESF with gem5 to take advantage of its processor simulation and to enable the end-to-end evaluation of real-world applications. Specifically, the gem5 memory system contains three major layers (\ie cache, memory controller, and underlying memory). Among the three layers, the memory controller performs as an interconnect level similar to the level of CXL protocol, which passes the memory accesses from caches to the underlying memory. It also manages different types of memories and provides a general view of memory to the CPU. 
To cooperate with gem5's native memory system, we extends gem5 MemCtrl with the interfaces to ESF to add CXL interconnection into the simulation. Specifically, as shown in Figure~\ref{fig:integrate}a, we implemented a Wrapper object, which utilizes the memory management functions of the MemCtrl. Each wrapper is integrated with two ESF devices, namely \emph{UpInterface} and \emph{DownInterface}. When a memory packet arrives, it is firstly passed to the UpInterface. The interface then transforms it into a packet and passes it to the DownInterface through a standalone ESF simulation to simulate the additional latency of the CXL system. This procedure is performed by reusing the gem5 event engine (\ie registering and simulating a set of gem5-style events). Upon the arrival of the packet at DownInterface, it will be transformed back to the gem5 memory packet, and the functions of the original gem5 MemCtrl are conducted. After the procedure in the underlying memory, the packet is passed back from DownInterface to UpInterface to simulate the response procedure.
One of the advantages of this implementation is that the wrapper can utilize the underlying memory objects from the original gem5, which enhances its extensibility. To support DMC functionality, we also implement a coherency interface using gem5's native tool SLICC. When the DCOH in ESF issues a back-invalidation request, it will be forwarded to the CohInterface. The interface will use gem5 native events to invalidate corresponding cachelines in the cache hierarchy to simulate the DMC function.

We also integrate ESF with two representative memory and storage simulators, namely \emph{DRAMsim3}~\cite{(DRAMsim3)Li2020DRAMsim3AC} and \emph{SimpleSSD}~\cite{(simplessd)Gouk2018AmberEP}. These simulators provide architectural details of various types of endpoints, including DDRx/HBM DRAM and SSD, which may exist in future CXL systems. In particular, we implemented wrappers for these simulators. For cycle-based simulator (\ie DRAMsim3), the wrapper periodically register a clocking event to make progress in DRAMsim3 simulation. For event-driven simulator (\ie SimpleSSD), the wrapper transforms the event format and registers them in the ESF event engine.
In summary, ESF is capable of evaluating CXL-based processors and endpoints by leveraging the existing simulators. 

Table~\ref{tab:othersim} summarizes the differences between ESF and several related simulation frameworks that simulate computing systems based on either CXL protocol or traditional interconnects. Specifically, most of the existing network-extended computation-centric simulation frameworks (\eg gem5-garnet~\cite{(garnet)Agarwal2009GARNETAD, (garnet3)Bharadwaj2020KiteAF} and BZSim~\cite{(BZSim)Strikos2024BZSimFL}) are mainly designed for on-chip interconnection simulation and lack detailed simulation for off-chip interconnection, such as the PCIe bus used by CXL. On the other hand, existing CXL-oriented simulators (\eg CXLMemSim~\cite{(CXLMemSim)Yang2023CXLMemSimAP} and MQSim\_CXL~\cite{(CXL-SSD)Yang2023OvercomingTM}) are constrained to simulate only one specific type of CXL devices (\ie memory for CXLMemSim and SSD for MQSim\_CXL) rather than a general computing system that have various types of CXL devices attached. In contrast, ESF can simulate the off-chip device interconnection, support diverse peripheral devices in CXL systems, and integrate with commonly-used simulators (\eg gem5, DRAMsim3, and SimpleSSD) for strong extensibility. ESF is provided as a unified research wheel for researchers who are interested in different fields, including architecture, distributed systems, and networks.

\section{Validation}
\label{sec:validation}
\begin{figure}[t]
\begin{minipage}{1\linewidth}
    \centering
    \includegraphics[width=1\linewidth]{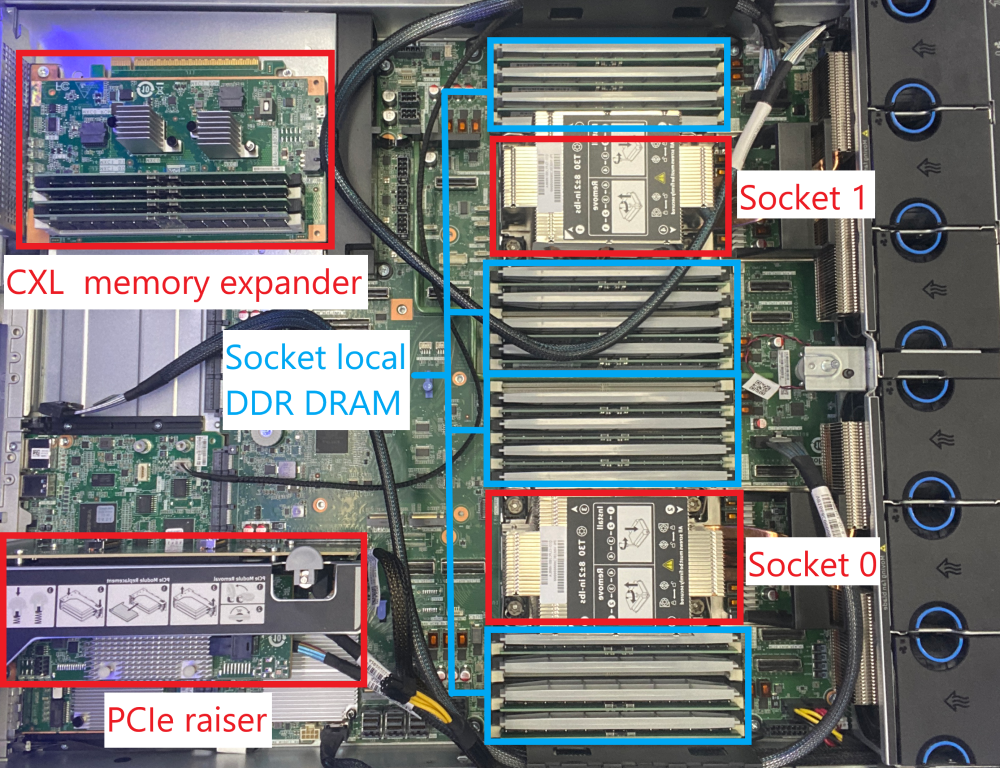}
    \caption{Top view of the hardware platform.}
    \vspace{5pt}
    \label{fig:hardware}
\end{minipage}
\begin{minipage}{1\linewidth}
    \centering
    \resizebox{1\linewidth}{!}{
    \begin{tabular}{|lr|lr|}
    \hline
        Requester process time & $10ns$ & PCIe port delay & $25ns$ \\
        Cache access time & $12ns$ & Bus time & $1ns$ \\
        Device controller process time & $40ns$ & Switching time & $20ns$ \\
    \hline
    \end{tabular}}
    \vspace{-5pt}
    \captionof{table}{Latency configurations of  critical components in validation.}
    \vspace{-25pt}
    \label{tab:lat}
\end{minipage}
\end{figure}

\noindent \textbf{Methodology.}
We validate ESF using a dual-socket platform with commercially available CXL hardware. The top view of our platform is shown in Figure~\ref{fig:hardware} (the CXL memory expander is unplugged for a better view). Each socket is equipped with an Intel Xeon Gold 6416H CPU~\cite{(Xeon)} and eight DDR5-4800 DRAM DIMMs, providing 128GB of main memory. One of the sockets is attached by a CXL memory expander with a CXL memory expander controller (MXC) from Montage Technology~\cite{(CXLMXC)}, supporting up to CXL 2.0 protocol and PCIe 5.0 $\times$16 standard. The memory expander consists of four DDR5-4800 DRAM DIMMs, providing 64GB of CXL HDM-H memory. To simulate the CXL system, we construct a sample system in ESF, which includes a requester, an interconnect bus, and four memory endpoints. For a fair comparison, we adjust the number of DRAM DIMMs in each NUMA node to four, which is equal to the DIMM number in the CXL memory expander.
In ESF we use the integrated DRAMsim3~\cite{(DRAMsim3)Li2020DRAMsim3AC} as the default endpoint components for an accurate DRAM simulation. For calibration, we follow the statistics measured on real hardware platforms and those from multiple prior works~\cite{(stats1)Sharma2022ComputeEL,(stats2)Gouk2022DirectAH,(stats3),(stats4),(Pond)Li2022PondCM,(TPP)Maruf2022TPPTP,(demystify)Sun2023DemystifyingCM}, and configure the components in ESF with these latency statistics. The detailed configurations are listed in Table~\ref{tab:lat}. For validation, we measure three major metrics: idle latency, peak bandwidth under different read-write ratio, and loaded-latency under different request intensity. To measure these metrics on hardware platform, we adopt Intel Memory Latency Checker (MLC)~\cite{(mlc)}, a widely used tool. The MLC is designed to evaluate a certain memory region for its idle latency and peak bandwidth. In addition, the read-write ratio of the requests generated by MLC is configurable. It can also run in loaded-latency mode, which varies the request intensity to measure the change of latency and bandwidth under different system loads. We run the tests of idle latency, peak bandwidth and loaded-latency with MLC based on CXL MXC memory and NUMA remote memory respectively to extract the metrics of CXL hardware and NUMA-emulator. For simulation platform, we directly generate memory accesses from the requester component to measure these metrics. By adjusting the queue size and issuing delay between requests, we can modify the request intensity to evaluate the loaded-latency. The idle latency and peak bandwidth are measured under fixed low and high system loads, respectively. The total amount of generated requests during each simulation test is 16000, and each endpoint receives 4000 requests. To warm up the system to the steady-state, 16000 additional requests are performed before collecting the final results. We also compare the simulation accuracy and speed of ESF (both standalone and gem-integration modes) with other platforms by running two example workloads from SPEC CPU2017~\cite{(spec17)Bucek2018SPECCN}. The cache hierarchy is configured to match our hardware platform (\ie 1.7MB L1D cache, 72MB L2 cache and 96MB L3 cache). For hardware platforms, the workloads are directly run by specifying the used socket and memory with \texttt{numactl}. For standalone mode, the memory access traces of the workloads are firstly collected with Intel PIN~\cite{(pin)} and filtered with a simulated cache hierarchy, then passed to ESF. For gem5-integrated simulators (\ie gem5-ESF and gem5-garnet), the workloads are run in gem5 SE mode.


\begin{figure}[t]
\vspace{5pt}
\begin{minipage}{1\linewidth}
    \centering
    \includegraphics[width=1\linewidth]{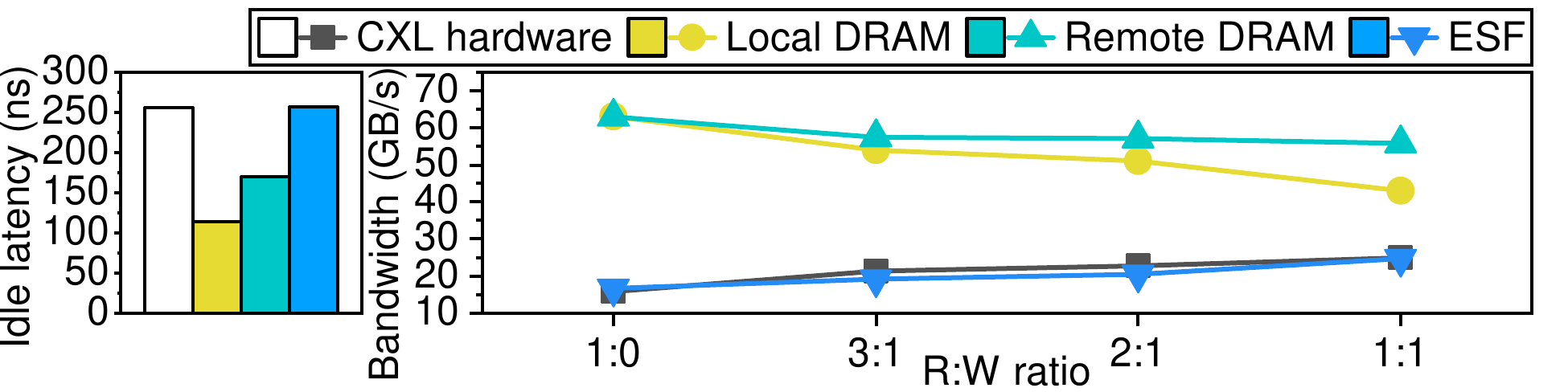}
    \vspace{-15pt}
    \caption{Idle latency and bandwidth of different platforms.}
    \label{fig:vali1}
\end{minipage}
\vspace{5pt}
\begin{minipage}{1\linewidth}
    \centering
    \includegraphics[width=1\linewidth]{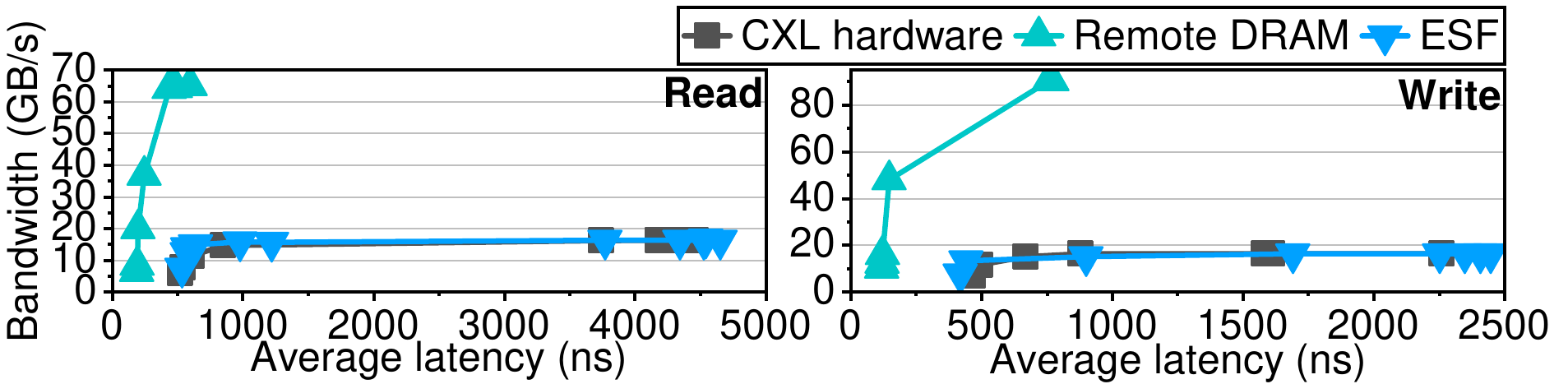}
    \vspace{-15pt}
    \caption{Latency-bandwidth curves of different platforms.}
    \vspace{0pt}
    \label{fig:vali_lb}
\end{minipage}
\begin{minipage}{1\linewidth}
\centering

\begin{tabular}{|c|cc|}
\hline
\multirow{2}{*}{Compared platforms} & \multicolumn{2}{c|}{SpecCPU2017 workloads} \\ \cline{2-3}
& \multicolumn{1}{c|}{gcc} & \multicolumn{1}{c|}{mcf} \\ \hline
\makecell[c]{CXL Hardware} & 18.0\% (0\%) & 24.2\% (0\%) \\
\makecell[c]{ESF standalone} & 18.7\%(+0.7\%) & 29.8\% (+5.6\%) \\
\makecell[c]{gem5-ESF} & 15.6\% (-2.4\%) & 19.8\% (-4.4\%) \\
\makecell[c]{NUMA emulation} & 20.0\% (+2.0\%) & 15.0\% (-9.2\%) \\
\makecell[c]{gem5-garnet} & 12.2\% (-5.8\%) & 15.2\% (-9.0\%) \\
\hline
\end{tabular}
\vspace{-5pt}
\captionof{table}{Simulated execution time overhead incurred by CXL memory of applications on different platforms.}
\label{tab:gem5}
\end{minipage}

\begin{minipage}{1\linewidth}
\centering
\begin{tabular}{|c|cc|}
\hline
\multirow{2}{*}{Workloads} & \multicolumn{2}{c|}{Compared platforms} \\ \cline{2-3}
     & \multicolumn{1}{c|}{gem5-ESF} & \multicolumn{1}{c|}{gem5-garnet} \\ \hline
 gcc &  1.7\% & 21.5\% \\
 mcf &  2.7\% & 24.5\% \\
 \hline
\end{tabular}
\vspace{-5pt}
\captionof{table}{Simulation time overhead incurred to vanilla gem5.}
\vspace{-20pt}
\label{tab:speed}
\end{minipage}
\end{figure}

\noindent \textbf{Results.}
Figure~\ref{fig:vali1} presents the results of idle latency and bandwidth. In addition to the CXL hardware and ESF, we also demonstrate the results of local DRAM and remote DRAM. As can be observed, after calibration, ESF exhibits an outstanding latency accuracy compared to NUMA-based emulators using remote DRAM. Regarding bandwidth, ESF shows acceptable errors ranging from 0.1\% to 10\% when compared to CXL hardware, while the remote DRAM modules do not accurately reflect the absolute value of CXL hardware. We also derive an observation that, as the read-write mixed ratio increases, the bandwidth of CXL hardware increases synchronously, which is well captured by ESF. In contrast, the results of local and remote DRAM show a decreasing trend in bandwidth. Further investigation on this characteristic can be found in Section~\ref{sec:exploration}. 

We also conduct loaded-latency tests on different platforms by adjusting the requester intensity. The results are shown as latency-bandwidth curves in Figure~\ref{fig:vali_lb}. The curves of ESF can accurately align with those of CXL hardware for both read and write requests, with an error margin up to only 12\%, and an average error of 4.3\%. In both low and high-intensity scenarios, ESF closely reflects the average latency observed on the CXL hardware platform. In contrast, the NUMA-based emulator presents curves that are completely apart from those of CXL hardware.

Table~\ref{tab:gem5} demonstrates the accuracy of different platforms on SpecCPU2017 workloads. Since the performance of real-world applications is highly related to the exact micro-architecture of hardware CPUs, which is unknown and cannot be accurately simulated, we, instead, use the execution time overheads incurred by CXL memory as the metric. This approach excludes the influence of CPU micro-architecture, allowing us to concentrate only on the memory systems. As observed, both ESF-standalone and gem5-ESF can accurately reflect CXL overhead in real-world workloads, with errors as low as 0.7\% compared to hardware results. However, NUMA-emulation and gem5-garnet demonstrate errors up to 9.2\% and 9.0\%, respectively. In summary, ESF exhibits surpassing simulation accuracy compared to the prior approaches.

We also compare the simulation speed of ESF with a representative simulator (\ie gem5-garnet). For a fair comparison, we compare gem5-garnet with the integrated mode of ESF (\ie gem5-ESF) to exclude the influence of gem5. According to the results shown in Table~\ref{tab:speed}, ESF only increases simulation time by 2\% compared to vanilla gem5 on average, while garnet incurs 22.5\% extra simulation time on average. The results indicate that ESF has a faster simulation speed than garnet.

\section{Design Space Exploration}
\label{sec:exploration}
ESF is designed to help researchers make deep observations in features of newest CXL that traditional tools struggle to support, including PBR, DMC and full-duplex transfer. In this section, we explore the performance impacts of these features across various system setups. All the following experiments perform warming-up requests to prepare the systems to steady-states, and only collect results under the steady-states.

\subsection{Impact of Different System Topologies\label{sec:5A}}
\begin{figure}[t]
    \centering
    \includegraphics[width=1\linewidth]{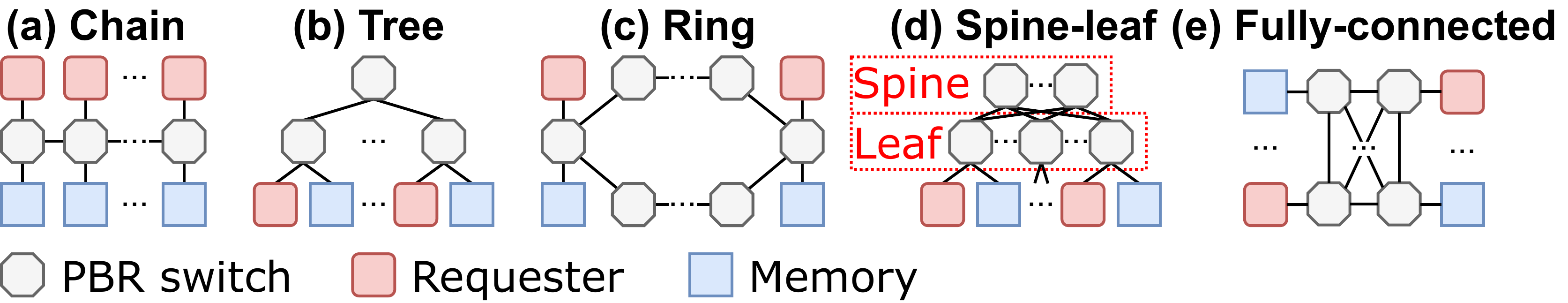}
    \vspace{-15pt}
    \caption{Examples of different system topologies.}
    \vspace{-15pt}
    \label{fig:topo_example}
\end{figure}

As discussed in Section~\ref{sec:2B}, the CXL protocol introduces PBR to support high scalability, which allows non-tree system topologies. To understand the impact on performance of system topologies, we perform a set of experiments using systems with $N$ requesters (\ie hosts and accelerators) and $N$ memory devices. The setup of a $N$-$N$ system is denoted as ``system scale = $2N$". In these experiments, the requesters issue random memory requests to all the memory devices. Different requesters and memory devices are connected via multiple PBR switches with different topologies. The bandwidth of a PBR switch port is constrained to a constant value. We investigated five types of topologies: (1) \texttt{chain}, (2) \texttt{tree}, (3) \texttt{ring}, (4) \texttt{spine-leaf} (SL), and (5) \texttt{fully-connected} (FC). Figure~\ref{fig:topo_example} shows the example diagrams of these topologies. 
During the simulation, each requester generates 4000 accesses to each endpoint, and the total request amount is $4000*N^2$.

\begin{figure}[t]
    \centering
    \includegraphics[width=1\linewidth]{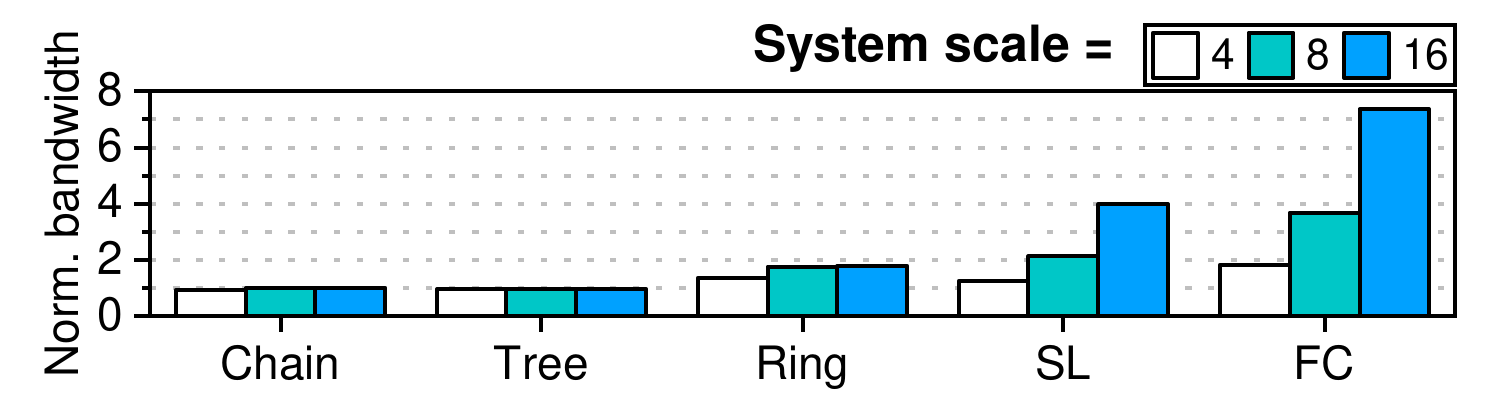}
    \vspace{-20pt}
    \caption{System bandwidth of different system topologies and scales, normalized to the max bandwidth of switch port.}
    \vspace{-20pt}
    \label{fig:swtopo_bw}
\end{figure}

\begin{figure}[t]
\begin{minipage}{1\linewidth}
    \centering
    \includegraphics[width=1\linewidth]{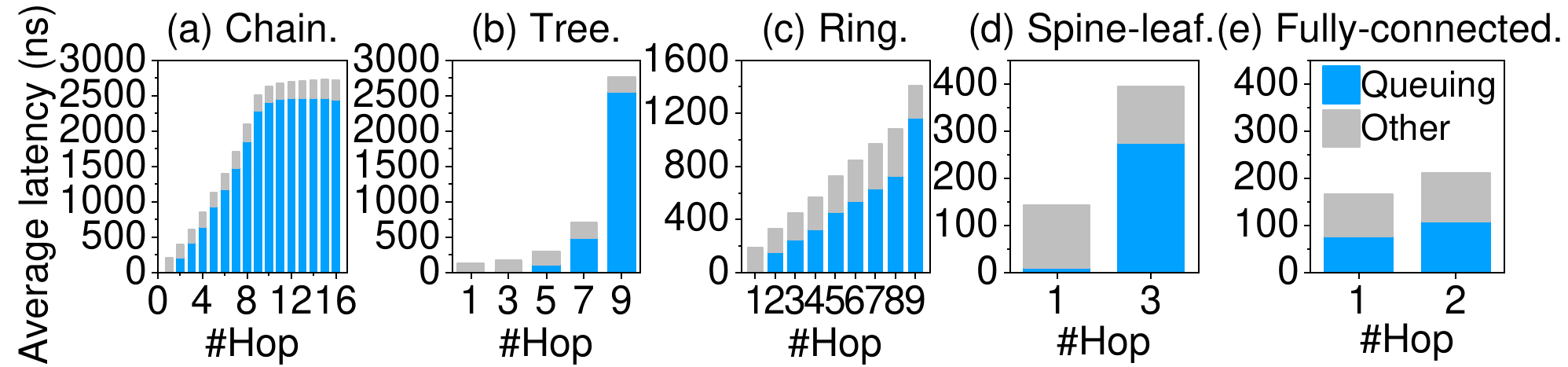}
    \vspace{-15pt}
    \caption{Latency of different system topologies, grouped by hop counts. The system scale is 16.}
    \label{fig:swtopo_lat}
\end{minipage}
\begin{minipage}{1\linewidth}
    \centering
    \includegraphics[width=1\linewidth]{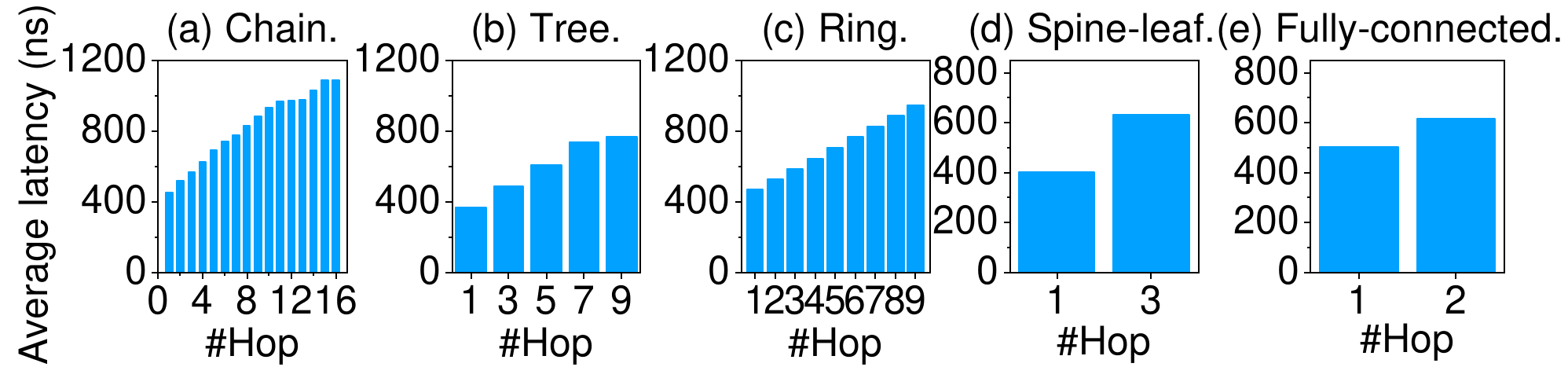}
    \vspace{-15pt}
    \caption{Latency of different system topologies under iso-bisection bandwidth condition, grouped by hop counts. The system scale is 16.}
    \vspace{-20pt}
    \label{fig:swtopo_bisec}
\end{minipage}
\end{figure}

\begin{figure}[t]
    \centering
    \includegraphics[width=1\linewidth]{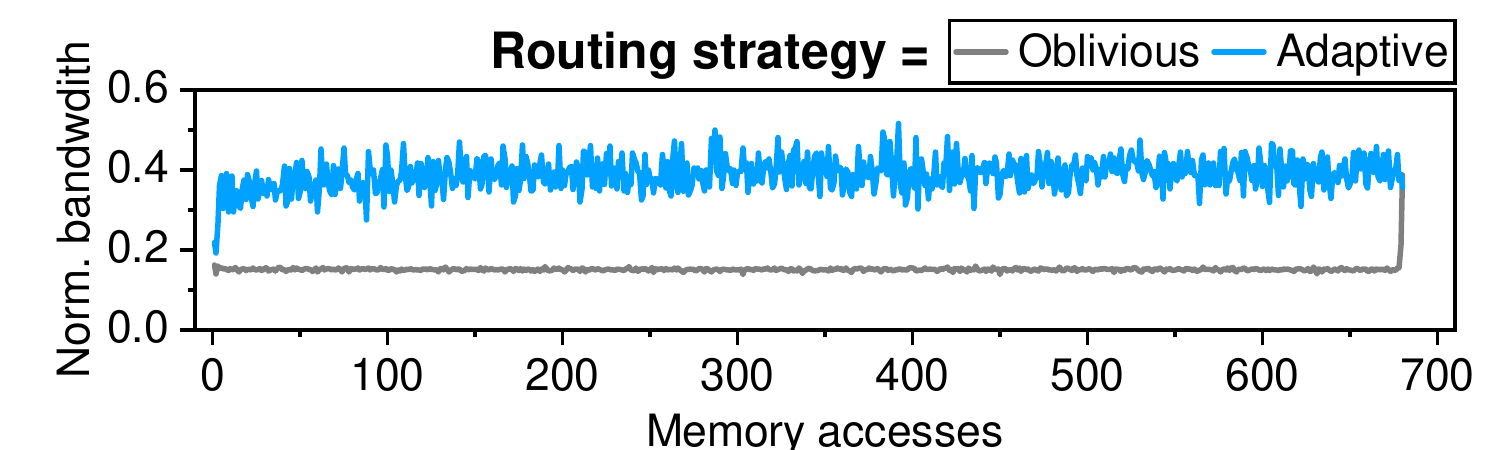}
    \vspace{-15pt}
    \caption{Bandwidth of observed host with different routing strategies, normalized to the max bandwidth of switch port.}
    \vspace{-20pt}
    \label{fig:route}
\end{figure}

\begin{figure*}
\begin{minipage}{0.49\textwidth}
    \centering
    \includegraphics[width=1\linewidth]{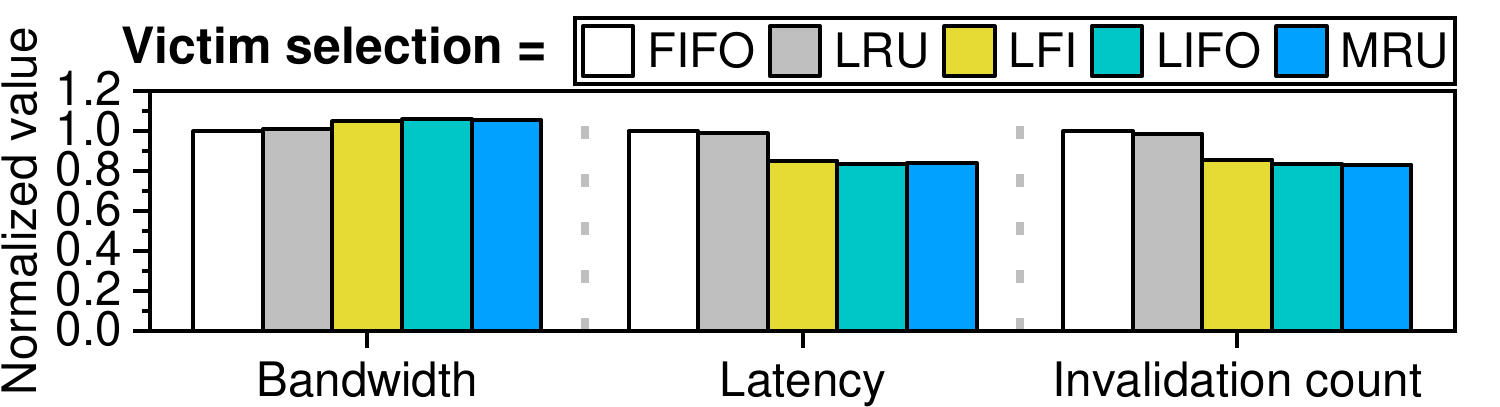}
    \vspace{-15pt}
    \caption{Performance of different snoop filter victim selections, normalized to \texttt{FIFO}.}
    \vspace{-15pt}
    \label{fig:thrash_hot}
\end{minipage}
\hspace{5pt}
\begin{minipage}{0.49\textwidth}
    \centering
    \includegraphics[width=1\linewidth]{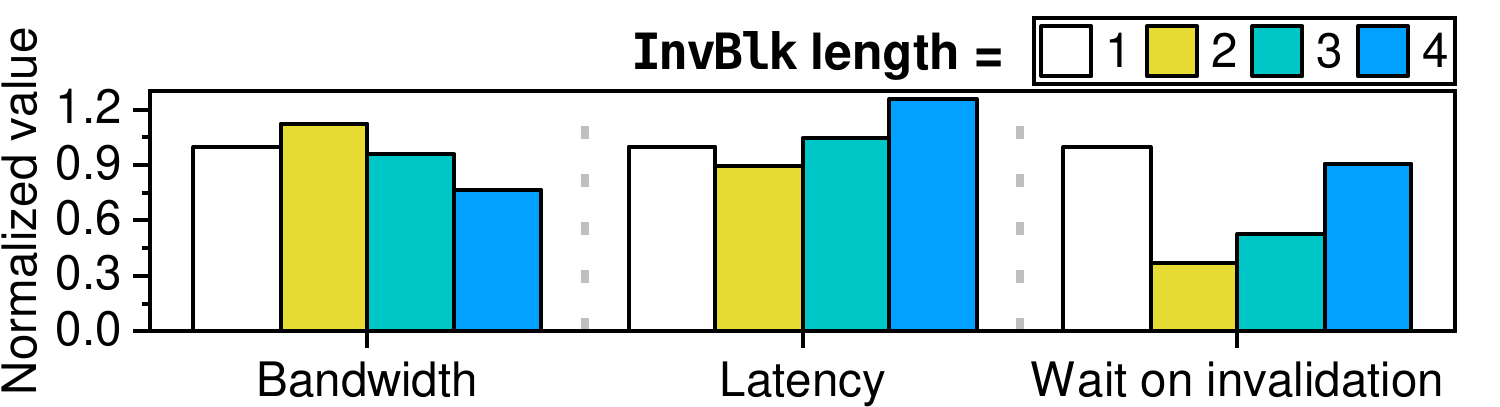}
    \vspace{-15pt}
    \caption{Performance of different \texttt{InvBlk} lengths, normalized to length=1.}
    \vspace{-15pt}
    \label{fig:invblk}
\end{minipage}
\end{figure*}
Figure~\ref{fig:swtopo_bw} illustrates the aggregated bandwidth in different systems. The bandwidth values are normalized to the maximum port bandwidth. The results highlight the bandwidth bottlenecks in different topologies. Both the \texttt{chain} and \texttt{tree} include ``bridge" routes (\ie all routes between switches in \texttt{chain} and routes directly connected to the root switch in \texttt{tree}), which are shared by all the requesters, limiting the system bandwidth to the maximum capacity of a single switch port. Scaling up the system with these topologies cannot improve the performance. The \texttt{ring} can provide an extra route in addition to \texttt{chain} and \texttt{tree}. Thus, by scaling up the system, the bandwidth can reach 2$\times$ of the port capacity. Compared to the former topologies, \texttt{spine-leaf} and \texttt{fully-connected} exhibit high scalability. The \texttt{spine-leaf} achieves this by replacing the bottleneck routes with a high-performance interconnect network (\ie the ``spine"). However, the competition among requesters on ports in ``leaf'' switches still exists. Thus, the \texttt{spine-leaf} can only provide $\frac{N}{2}\times$ bandwidth of the port capacity. The \texttt{fully-connected} overcomes this limitation with a network where each pair of devices can communicate directly. As a result, each requester is provided with full port bandwidth, achieving a system bandwidth of $N\times$ port capacity.

Figure~\ref{fig:swtopo_lat} depicts the average latency of requests across various topologies with a system scale of 16. The results are grouped by the number of hops the requests experienced. It can be observed that as the number of hops increases, the request will experience more switch queuing time, harming the performance. The bottleneck of ``bridge" routes also significantly impacts the latency. For example, as shown in Figure~\ref{fig:swtopo_lat}b, the queuing time of requests with 9 hops is drastically higher than that of other requests. This is because these requests are congested on the ``bridge" routes. In contrast, when being provided with an extra route in \texttt{ring}, the maximum latency is halved compared to the maximum latency in \texttt{tree} and \texttt{chain}, and the variation among latency with different hop counts becomes more balanced. Latency degradation due to congestion can also be found in \texttt{spine-leaf} system, where requests may congest at leaf switches and experience queuing time increasing. We also conduct ISO-bisection bandwidth tests on different system topologies, which are configured to deliver the same bisection bandwidth. The average latency results are shown in Figure~\ref{fig:swtopo_bisec}. Since the maximum port bandwidths of different system topologies can vary under ISO-bisection bandwidth configuration, which leads to unmatched switch queuing time, we only show the overall average latency values of different hop numbers and omit the breakdown. Although the average latency values of \texttt{chain}, \texttt{tree} and \texttt{ring} are decreased, increasing the hop number can still cause congestion on critical paths, incur significant latency overhead (\ie about 2$\times$ in \texttt{chain} and 1$\times$ in \texttt{tree} and \texttt{ring} compared to the latency with lowest hop number) and introduce latency unpredictability. In contrast, since \texttt{spine-leaf} and \texttt{fully-connected} require fewer hops due to their specific network structures, they can provide high stability for latency values, and achieve high system scalability. These results indicate that the traditional tree-like topology experiences significant system performance bottlenecks.

We then investigate the impact of routing strategy under a typical high-performance and low-cost system topology (\ie \texttt{spine-leaf}). The routing strategy is categorized into two classes, namely \texttt{Oblivious} and \texttt{Adaptive}~\cite{(route1)Trik2022AHS,(route2)Singh2022ReviewAA}. \texttt{Oblivious} routes every packet statically based on the source and destination, while \texttt{Adaptive} chooses the next hop dynamically according to the congestion condition among the available choices. The system is configured to include eight memory endpoints, eight noisy neighbors that intensively access the memories, and a host that accesses the memories at a fixed rate. Figure~\ref{fig:route} shows the observed bandwidth of the host, normalized to the maximum port bandwidth. The results show that \texttt{Adaptive} routing strategies, compared to \texttt{Oblivious}, can drastically improve host performance under the pressure of noisy neighbors, indicating the importance of the routing strategy.


\begin{figure*}
\begin{minipage}{0.49\textwidth}
    \centering
    \includegraphics[width=1\linewidth]{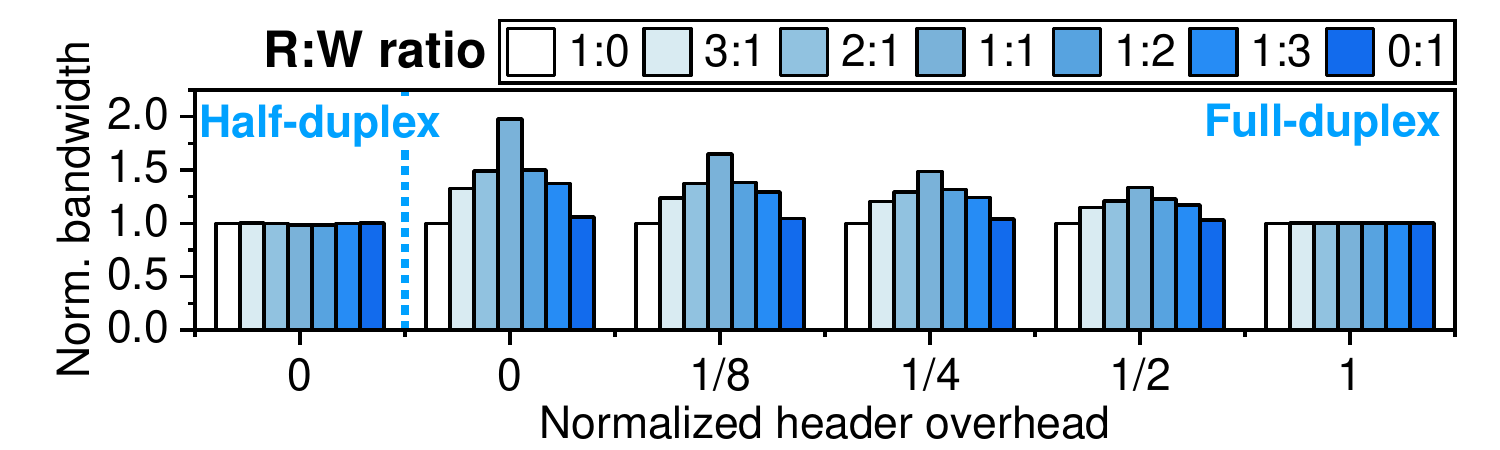}
    \vspace{-15pt}
    \caption{Bandwidth under different R:W ratios and header overheads, normalized to read-only scenarios.}
    \vspace{-20pt}
    \label{fig:rwmix_bw}
\end{minipage}
\hspace{5pt}
\begin{minipage}{0.49\textwidth}
    \centering
    \includegraphics[width=1\linewidth]{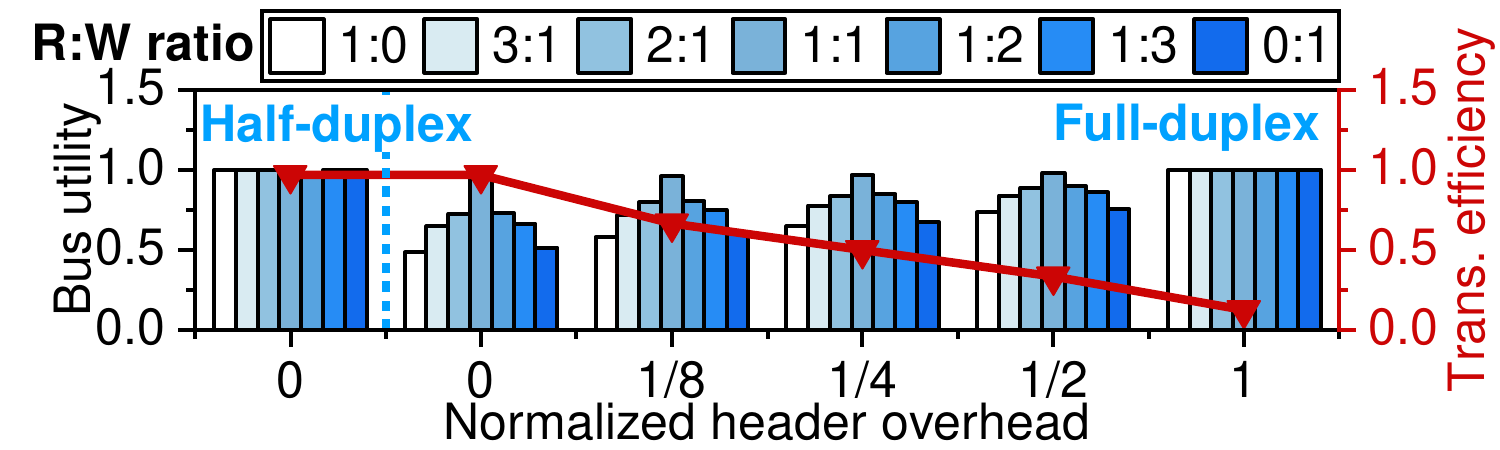}
    \vspace{-15pt}
    \caption{Bus utility and transmission efficiency under different R:W ratios and header overheads.}
    \vspace{-20pt}
    \label{fig:rwmix_util}
\end{minipage}
\end{figure*}

\subsection{Impact of Snoop Filter Victim Selection Policies\label{sec:5B}}
As mentioned in Section~\ref{sec:background}, the CXL protocol requires devices with DMC to implement an inclusive snoop filter (SF). Due to its inclusive nature, in cases of insufficient buffer entries, the SF will acquire new clear entries by selecting victims and issuing back-invalidate snoop (BISnp) requests to their current owners. These BISnp requests will clear the lines in the owners' local cache, impacting the system performance. Therefore, a goal of SF victim selection policy is to reduce the frequency of BISnp. 

To investigate the impact of different victim selection policies in SF, we test five basic policies. 
The tested system includes a requester, which issues coherent requests in a skewed pattern with 90\% to hot data and 10\% to cold data. The amount of hot data takes 10\% of the total memory footprint. The requester is equipped with a local cache that filters the hit events. The size of the cache is configured to 20\% of the total memory footprint, making sure it can cache all the hot data. In cases of a cache miss, the request is routed to the memory device through a bus, which is configured with infinite bandwidth to eliminate unexpected performance impact. On the device side, an SF filters the requests and issues BISnp whenever necessary. A BISnp will be sent to the requester to invalidate the corresponding cacheline. The size of the SF is set to be the same with the local cache size in order to record states of all cached data. 
We test the following victim selection policies: (1) \texttt{FIFO} (First-In, First-Out), (2) \texttt{LRU} (Least Recently Used), (3) \texttt{LFI} (Least Frequently Inserted), (4) \texttt{LIFO} (Last-In, First-Out), and (5) \texttt{MRU} (Most Recently Used). 
The number of endpoints is set to four, and each endpoint receives 4000 accesses during the evaluation.

Figure~\ref{fig:thrash_hot} depicts the results of bandwidth, latency, and invalidation count, all normalized to \texttt{FIFO}. Since there is little hit event in the SF, \texttt{FIFO} and \texttt{LRU} behave similarly to \texttt{LIFO} and \texttt{MRU}, respectively. Compared to \texttt{FIFO}, \texttt{LIFO} improves the bandwidth by 5\%, while decreasing the average latency and invalidation count by 15\% and 16\%, respectively. The difference between the SF and local cache explains these findings. As the system reaches its steady state, most of the hot data reside in the local cache, while the SF records the coherence states of these hot data. Most of the requests reaching the SF are cache misses, targeting cold data. In this scenario, the ``last-in" or ``most recent" entries, rather than the ``first-in" or ``least recent" entries, actually store information for cold data and are the suitable victims. In contrast, the \texttt{FIFO} and \texttt{LRU} are more likely to invalidate hot data, harming the system performance. To demonstrate the impact of invalidating hot data, we also propose and evaluate the \texttt{LFI} policy, which maintains a global counter table to record the inserted times of each cacheline. Upon invalidation, \texttt{LFI} selects the least frequently inserted address as the victim to avoid invalidating hot data. The results show that \texttt{LFI} reduces invalidation count by 15\% compared to \texttt{FIFO}, proving that \texttt{FIFO} invalidates hot data more frequently. However, since the \texttt{LFI} leverages global information, it will periodically invalidate all the hot cachelines when their inserted times become equal. This leads to a slightly worse performance compared to \texttt{LIFO} and \texttt{MRU}.


\subsection{Impact of \texttt{InvBlk} Commands}
The CXL protocol proposes a set of \texttt{InvBlk} commands for the SF. When the SF sends a BISnp request, it can additionally send a \texttt{InvBlk} command, which requires the owners to invalidate a sequence of cachelines with contiguous addresses. The length of these cachelines can range from two to four. This feature is introduced to improve the efficiency of BISnp, allowing the SF to clear multiple entries with a single request. To understand the impact of \texttt{InvBlk} commands, we perform a set of experiments on a system with two requesters issuing sequential requests, a local cache in each requester, a bus, and a memory device with an SF. The configurations including cache size, SF size and request number are the same with those described in Section~\ref{sec:5B}. To zoom-in the effects of \texttt{InvBlk} commands, the SF employs a block-length-prioritized victim selection policy. During victim selection, the SF chooses the longest sequence of entries with contiguous addresses. It leverages \texttt{LIFO} policy among multiple possible victims. In our experiments, we limit the maximum length of entry sequences to evaluate the impact of \texttt{InvBlk}.

Figure~\ref{fig:invblk} depicts the results of bandwidth, average latency, and average waiting time for invalidation. When the \texttt{InvBlk} length is larger than one, a single BISnp request can clear more than one entry. As a result, subsequent coherent requests no longer need to wait on BISnp, reducing the average waiting time. When two lines are cleared in one BISnp, this benefit brings the reduction of total average latency and the increase of bandwidth. However, when the number of lines in a BISnp exceeds two, the overhead of access requester local caches increases, diminishing the benefit of \texttt{InvBlk}. Furthermore, the data within the BISnp flows compete for the transmission bandwidth. As a result, the performance of larger \texttt{InvBlk} length shows no improvement compared to length=1. 

\begin{figure*}
\begin{minipage}{0.49\textwidth}
    \centering
    \includegraphics[width=1\linewidth]{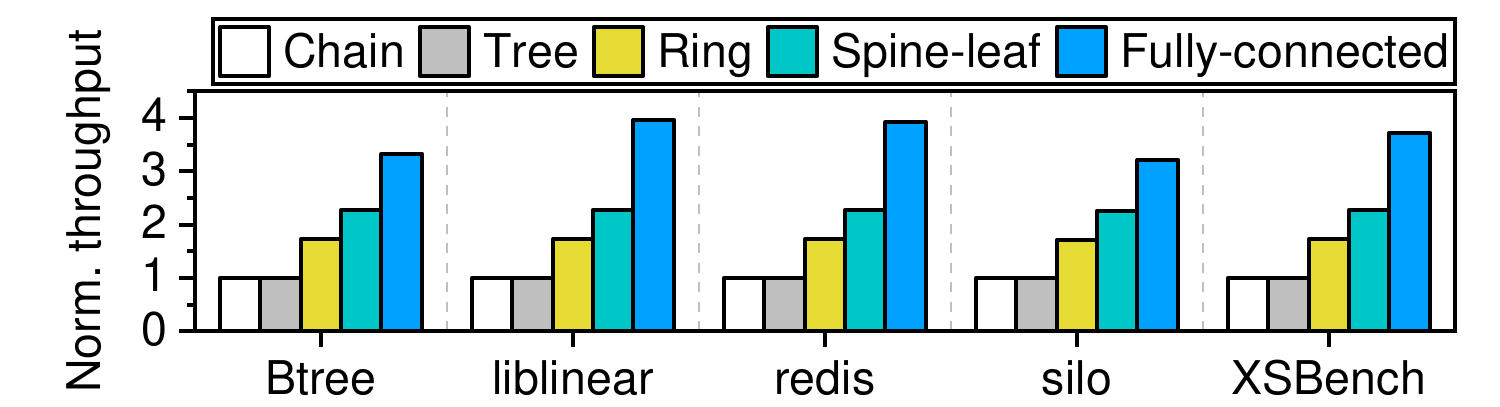}
    \vspace{-15pt}
    \caption{Throughput of different real-world traces with different system topologies, normalized to \texttt{Chain}.}
    \vspace{-20pt}
    \label{fig:trace_topo_bw}
\end{minipage}
\hspace{5pt}
\begin{minipage}{0.49\textwidth}
    \centering
    \includegraphics[width=1\linewidth]{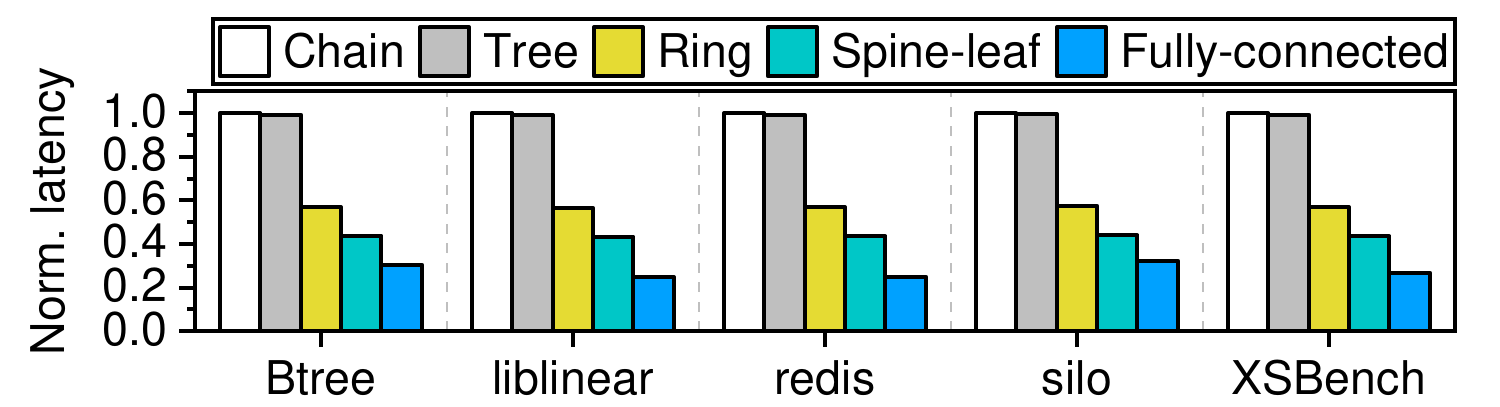}
    \vspace{-15pt}
    \caption{Average latency of different real-world traces with different system topologies, normalized to \texttt{Chain}.}
    \vspace{-20pt}
    \label{fig:trace_topo_lat}
\end{minipage}
\end{figure*}
\begin{figure}
    \centering
    \includegraphics[width=1\linewidth]{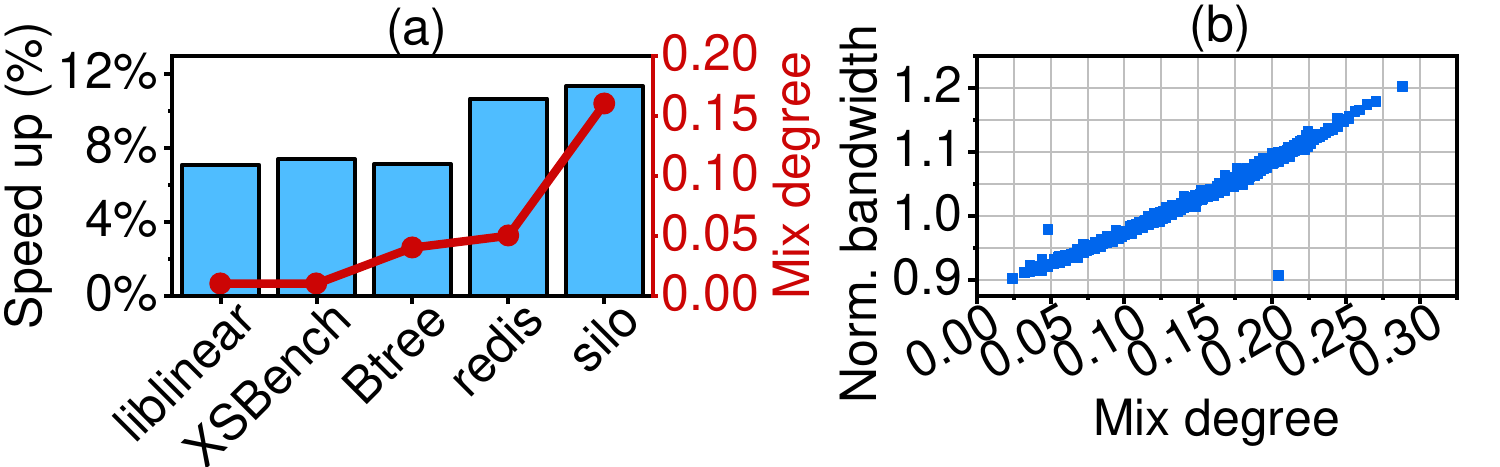}
    \vspace{-20pt}
    \caption{(a) Execution speedup of full-duplex bus against half-duplex bus and mix degrees of different real-world traces. (b) Performance of \texttt{silo} with full-duplex bus, normalized to the max bandwidth of one bus direction.}
    \vspace{-20pt}
    \label{fig:trace_bus}
\end{figure}

\subsection{Full Duplex Transmission\label{sec:5D}}
We then present our exploration with ESF on the impact of PCIe bus' full-duplex transfer feature.
We build a dedicated system to evaluate the effects of this feature, which includes a requester issuing random requests based on a read-write ratio setup, a bus incurring packet size overheads to the header packets, and four memory devices. Besides the bandwidth, we define two other metrics: (1) bus utility, indicating the fraction of time when the bus is busy compared to the total simulation time (average in all transmission directions), and (2) transmission efficiency, denoting the fraction of time the bus spends on payload transmission compared to total bus transmission time. Across experiments, we adjust the read-write ratio and the incurred header overheads to understand the impact of full-duplex transmission under different scenarios. The system includes four endpoints, each of them receives 4000 requests during the simulation.

Figure~\ref{fig:rwmix_bw} depicts the bandwidth results. The header overheads are normalized to payload length, and the bandwidth values are respectively normalized to read-only scenarios for each header overhead setting. The figure shows that, with all other configurations unchanged, the bandwidth of a full-duplex bus system is more affected by the read-write ratio than that of a half-duplex bus system. Specifically, the system bandwidth stays almost constant for a half-duplex bus. On the other hand, mixing read and write requests enhances the bandwidth of a full-duplex bus system. These findings are consistent with the hardware platform observations discussed in Section~\ref{sec:validation}. We also conduct the tests by varying the header overhead. As can be observed, with zero header overhead, a 1:1 mix of read and write packets can nearly double the system bandwidth. As header overhead increases, the improvement of read-write mixing decreases. When the headers have the same length of payloads, the improvement drops to zero. We explain this phenomenon as follows. In the case of full-duplex buses, when memory accesses are only with a single operation type, the packets utilize only one direction of the bus, leaving the opposite direction for zero-payload headers. Consequently, the opposite direction is not fully utilized. When the read and write packets are mixed, both directions of the bus are engaged in transmitting payloads, thereby improving the overall bandwidth. On the other hand, a half-duplex bus only provides a single direction at one time, thus, there are no space left for improving the system bandwidth.

To further explore on this phenomenon, we demonstrate the evaluation results on bus utility and transmission efficiency.  Figure~\ref{fig:rwmix_util} shows the results. In the case of half-duplex bus, the bus is almost fully utilized, and the utility remain unchanged as read-write ratio varies. On the other hand, the bus utility of full-duplex bus is heavily affected by read-write ratio. When the header overhead is zero, single-type scenario only utilized half of the bus (\ie one direction in a total of two). As the mix ratio increases, the bus utility approaches 1, which means both directions of the bus are fully utilized. These results indicate that the bandwidth improvement of read-write mix scenarios comes from the improved bus utility. By increasing the header overhead, more bus time will be cost on non-payload transmission, decreasing the transmission efficiency. Because more time is spent on header transmission, the bus utility of single-type scenarios increases, leaving less space for read-write mixing to improve utility. As a result, the bandwidth improvement is also weakened, as can be observed in Figure~\ref{fig:rwmix_bw}. These results support the explanation that, in scenarios with a single type of operation, the headers will waste one of the bus directions, and read-write mixed scenarios can optimize bus utility, leading to bandwidth improvement.




\subsection{Impact on Real-World Workloads}
We also investigate the impact of CXL systems on real-world workloads by replaying the traces of five representative workloads (\ie BTree~\cite{(mitosis)Achermann2019MitosisTS}, liblinear~\cite{(liblinear)fan2008liblinear}, redis~\cite{(redis),(ycsb)Cooper2010BenchmarkingCS}, silo~\cite{(silo)Tu2013SpeedyTI} and XSBench~\cite{(XSBench)Tramm:wy}) using ESF. The memory traces, each containing one million memory accesses, are collected by using a popular tool proposed by prior work~\cite{(CXL-SSD)Yang2023OvercomingTM}.

Figure~\ref{fig:trace_topo_bw} and~\ref{fig:trace_topo_lat} demonstrate the evaluated results of these workloads running on different system topologies, which are described in Section~\ref{sec:5A}. The system topology significantly impacts the performance of all real-world workloads. Similar to our aforementioned observation, both the \texttt{chain} and the \texttt{tree} exhibit the lowest throughput and highest average memory latency compared with other topologies. This is because of the bottleneck route in these traditional system topologies. By simply widening the system with an extra route, \texttt{ring} can achieve 1.72$\times$ throughput and 0.57$\times$ average latency, compared to \texttt{chain} and \texttt{tree}. The \texttt{spine-leaf} and \texttt{fully-connected} topologies achieve higher performance by eliminating the bottleneck routes, achieving 2.27$\times$, 3.63$\times$ throughput and 0.44$\times$, 0.28$\times$ average latency, respectively. These results confirm that system topology notably affects the performance of CXL memory system in terms of both throughput and latency, and the traditional tree-like topologies incur significant overheads.

Figure~\ref{fig:trace_bus}a and~\ref{fig:trace_bus}b depict the impact of PCIe full-duplex transfer on the real-world workloads. As discussed in Section~\ref{sec:5D}, mixing read and write requests can improve the system performance with a full-duplex bus by utilizing both transfer directions. This impact is also observed in real-world workloads. As shown in Figure~\ref{fig:trace_bus}a, when the mix degree (defined as $\min\{read\_ratio, write\_ratio\}$) of a workload increases, its speed-up compared to a half-duplex platform also increases. Figure~\ref{fig:trace_bus}b further demonstrates the relationship between the mix degree and performance. In the figure, each point represents the bandwidth of 1000 memory accesses, normalized to the max bandwidth of one bus direction. It can be observed that, there is a high-positive correlation between mix degree and performance. When the mix degree increases by 0.1, the overall bandwidth can be improved by 9\%. This observation suggests that real-world workloads can mix memory read and write more aggressively when running on CXL memory to achieve better performance.


\section{Conclusion}
\label{sec:conclusion}
In this work, we introduce ESF, a novel simulation framework that accurately models several critical features in CXL 3.1 specification, that existing emulation and simulation tools struggle to support. These features help ESF to simulate CXL systems with high scalability and coherent peer-to-peer communication. We validate ESF on a real CXL-attached hardware platform and demonstrate outstanding accuracy compared to emulators adopted by prior works. ESF can uncover important issues that existing tools cannot figure out, such as the performance impact of device-managed coherence. We hope ESF can assist in the exploration of high-performance CXL system design.


\bibliographystyle{IEEEtranS}
\bibliography{refs}

\end{document}